\RequirePackage{fix-cm}
\documentclass[smallextended]{svjour3}       
\usepackage{natbib}

\smartqed  
\usepackage{graphicx}
\journalname{Empirical Software Engineering}
\usepackage{amsmath,amssymb,amsfonts}
\usepackage{algorithmic}
\usepackage{textcomp}
\usepackage{rotating}
\usepackage{multirow}
\usepackage[dvipsnames]{xcolor}
\usepackage{tikz}
\usepackage{pgf,pgfplots}
\pgfplotsset{compat=1.14}
\usetikzlibrary{fit,calc,pgfplots.statistics,positioning,shapes.geometric,shapes.arrows,shapes.symbols,svg.path,patterns,intersections,fillbetween,decorations.text,shapes,trees,arrows.meta}
\usepgfplotslibrary{external}
\tikzstyle{commit} = [circle, inner sep=0mm, outer sep=0,draw,minimum width=0.5cm]
\tikzset{
    fontscale/.style = {%
    font=\relsize{#1}
    },
    arrow/.style = {%
    draw=blue!30,
    line width=2mm,
    -{Triangle[length=3mm]},
    shorten >=1mm, shorten <=1mm,
    font=\fontsize{8}{8}\selectfont
    }
}
\usepackage{caption,booktabs}
\usepackage{hyperref}

\newcommand{\numberProjects}{54}
\newcommand{\numberCommits}{112,266}  

\makeatletter
\newbox\dottedarrow@box
\setbox\dottedarrow@box\hbox
  {%
    \begin{tikzpicture}
      \draw[dotted,->] (0,0) -- (1.5em,0);
    \end{tikzpicture}%
  }
\newcommand*\dottedarrow
  {\relax\ifmmode\expandafter\dottedarrow@m\else\expandafter\dottedarrow@t\fi}
\newcommand*\dottedarrow@t[1][1.5em]
  {\resizebox{#1}{!}{\raisebox{.5ex}{\usebox\dottedarrow@box}}}
\newcommand*\dottedarrow@m[1][]
  {%
    \if\relax\detokenize{#1}\relax
      \mathchoice
        {\dottedarrow@t}
        {\dottedarrow@t}
        {\dottedarrow@t[1.1em]}
        {\dottedarrow@t[0.9em]}%
    \else
      \dottedarrow@t[#1]%
    \fi
  }
\makeatother

\usepackage[title]{appendix}
\usepackage{tabularx}
\newcolumntype{L}{>{\raggedright\arraybackslash}X}
\usepackage{ltablex}

\usepackage[normalem]{ulem}  
\newcommand{\rev}[1]{#1}
\newcommand{\del}[1]{}
\begin{document}

\title{A Longitudinal Study of Static Analysis Warning Evolution and the Effects of PMD on Software Quality in Apache Open Source Projects}
\titlerunning{Static Analysis Warning Evolution and the Effects of PMD}        

\author{Alexander Trautsch \and
        Steffen Herbold \and
        Jens Grabowski
}

\institute{Alexander Trautsch\\Institute of Computer Science, University of Goettingen, Germany\\
           \email{alexander.trautsch@cs.uni-goettingen.de}
           \vspace{5pt}\\
           Steffen Herbold\\Institute of Computer Science, University of Goettingen, Germany\\
           \email{herbold@cs.uni-goettingen.de}
           \vspace{5pt}\\
           Jens Grabowski\\Institute of Computer Science, University of Goettingen, Germany\\
           \email{grabowski@cs.uni-goettingen.de}
}

\date{Received: date / Accepted: date}

\maketitle

\begin{abstract}
Automated static analysis tools (ASATs) have become a major part of the software development workflow.
Acting on the generated warnings, i.e., changing the code indicated in the warning, should be part of, at latest, the code review phase.
Despite this being a best practice in software development, there is still a lack of empirical research regarding the usage of ASATs in the wild.
In this work, we want to study ASAT warning trends in software via the example of PMD as an ASAT and its usage in open source projects. We analyzed the commit history of \numberProjects{} projects (with \numberCommits{} commits in total), taking into account 193 PMD rules and 61 PMD releases.
We investigate trends of ASAT warnings over up to 17 years for the selected study subjects regarding changes of warning types, short and long term impact of ASAT use, and changes in warning severities.
We found that large global changes in ASAT warnings are mostly due to coding style changes regarding braces and naming conventions.
We also found that, surprisingly, the influence of the presence of PMD in the build process of the project on warning removal trends for the number of warnings per lines of code is small and not statistically significant. 
Regardless, if we consider defect density as a proxy for external quality, we see a positive effect if PMD is present in the build configuration of our study subjects.
\end{abstract}

\section{Introduction}
Automated static analysis tools (ASATs) support software developers with warnings and information regarding common coding mistakes, design anti-patterns like code smells~\citep{fowler}, or code style violations. ASATs work directly on the source code or bytecode without executing the program.
They are using abstract models of the source code, e.g., the Abstract Syntax Tree (AST) or the control flow graph to match the provided source code against a set of rules defined in the ASAT. If a part of the source code violates a predefined rule, a warning is generated. These rules can be customized by the project using the ASAT to fit their needs by removing rules deemed unnecessary. ASAT reports usually contain a type of warning, a short description, and the file and line number of the source code that triggered the warning. Developers can then inspect the line specified in the warning and decide if a change is necessary.

The defects that can be found by static analysis include varying severities. Java String comparisons with ``=='' instead of using the equals() method, would compare the object reference instead of the object contents.
The severity rating for this type of warning is critical as it can lead to undesired behavior in the program.
Naming convention warnings, e.g., not using camel case for class names on the other hand have a minor severity.

ASATs are able to uncover problems with significant real world impact. The Apple Goto Fail defect~\footnote{https://www.imperialviolet.org/2014/02/22/applebug.html, last accessed: 2018-11-19} for example could have been detected by static analysis utilizing the control flow graph.
This importance regarding software quality is further demonstrated by the inclusion of ASATs in software quality models, e.g., Quamoco~\citep{quamoco} and ColumbusQM~\citep{columbus}.
\cite{Zheng2006} found that the number of ASAT warnings can be used to effectively identify problematic modules. Moreover, developers also believe that static analysis improves quality as reported by a survey of Microsoft employees by \cite{Devanbu2016}.

ASATs can be integrated as part of general static analysis via IDE plugins where the developer can see the warnings almost instantly. Usually IDE plugins are able to access a central configuration for rules that generate warnings. A central rule configuration is essential for project specific rules and exclusions of rules and directories.
Integrating ASATs in the software development process as part of the buildfile of the project has the advantage of providing a central point of configuration which can also be accessed by IDE plugins.
It also enables the developer to view generated reports prior to, or after the compilation as part of the build process.
Moreover, the inclusion into the buildfile also allows Continuous Integration (CI) Systems to generate reports automatically\del{ which}\rev{. The reports} can then be used to plot trends for general quality management or provide assistance in code reviews.
Published industry reports share some findings regarding static analysis infrastructure and warning removal.
Google~\citep{google_static} and Facebook~\citep{infer_scaling} both found that just presenting developers with a large list of findings rarely motivates them to fix all reported warnings\del{, whereas}\rev{. However,} reporting the warnings as early as possible, or at latest at code review time, improves the adoption and subsequent removal of static analysis warnings.
One of the lesson that Facebook and Google learned, is that static analysis warnings are not removed in bulk, but as part of a continuous process when code is added after the static analysis is configured or old code is changed.

ASAT warnings are able to indicate software quality because of how the rules that trigger the warnings are designed. 
The ASAT developers designed these rules not only to remove obvious bugs but also to express what is important for high quality source code via the designed rules.
Therefore, a lot of the existing rules are based on best practices, common coding mistakes and coding style recommendations.
Best practices and coding styles are also subject to evolution as the user base of a programming language evolves, new tooling is created and also as a programming language itself gets new features.
For example, the Java code written today is different than the Java code written 10 years ago.
These differences and, more importantly, the evolution of the usage of best practices are an interesting research topic, e.g., language feature evolution~\citep{python2to3}, and design pattern evolution~\citep{evolution_of_design_patterns}.

The topics covered in research with regards to ASATs are concerned with configuration changes~\citep{state_static}, CI-pipelines~\citep{ci_static}, finding reported defects~\citep{habib,thung,vetro2011} or warning resolution times~\citep{marcelline,debt_apache}.
\cite{Vassallo2019} provide a thorough investigation of developer usage of ASATs in different developing contexts.
One of the problems identified for ASAT usage is the number of false positives~\citep{whynot_static,Christakis2016,Vassallo2019} for which warning priorization~\citep{Kim2007,warnings_static} was proposed, sometimes as actionable warning identification~\citep{Heckman2009}, see also the systematic literature review by~\cite{Heckman2011}.
Developers perceive ASATs as quality improving~\citep{marcelline,Devanbu2016} although the percentage of resolved ASAT warnings vary, e.g., 0.5\%~\citep{tse_findbugs}, 8.77\%~\citep{marcelline}, 6\%-9\%~\citep{warnings_static} and 36.3\%~\citep{debt_apache}.

What is still missing, is a longitudinal, more general overview of the evolution of ASAT warnings over the years of development which includes complete measurement of ASAT warnings over the complete development history.
This would improve our understanding of exactly how the warnings evolve, e.g., how ASAT tools are used and the impact on the overall numbers of warnings over the project evolution.
Moreover, a direct linking between static analysis warnings and removal of the implicated code in the process of bug fixing may be limiting the insights that can be gained from investigating ASAT usage. 
Most ASATs are also detecting problems due to spacing, braces, readability, and best practices which are not directly causing a defect. Therefore, the influence of ASATs on defects or software quality as a whole may be more indirect.
To the best of our knowledge, only the work by~\cite{static_workshop} directly investigates this so far (there is also the work by~\cite{defect_prediction_findbugs} however it is not directly investigating correlations). Their work shows a positive correlation between ASAT warnings and defects, although their empirical study is limited to one project.

In this article, we investigate the usage of \rev{one} ASAT in \rev{Java} open source projects \rev{of the Apache Software Foundation} in the context of software evolution.
We determine the trends of removal of code with ASAT warnings over the projects lifetime. We are interested in the evolution of ASAT warnings on a project and on a global level, i.e., ASAT warnings for all projects combined. We examine general trends independent of developer interaction, i.e., is the state of software generally improving with regards to ASAT warnings. We also investigate which types of ASAT warnings have positive and negative trends to infer which types of coding standards or best practices are important to developers. To this end, we are not only interested in the absolute numbers of ASAT warnings but put them in relation to the project size. Moreover, we investigate the impact of including an ASAT in the build process on ASAT warning trends regarding their resolution.
Additionally, we approximate the impact of including an ASAT in the build process on external quality via defect density~\citep{fenton} by including defect information.

Our longitudinal, retrospective case study results in the following contributions of this work:
\begin{itemize}
    \item An analysis of evolutionary trends of ASAT warnings in \numberProjects{} open source projects from 2001-2017.
    \item An assessment of the effects of ASAT usage in open source projects on warning trends and software quality via defect density.
    \item An extension of prior work by providing a broader, long-term, evolutionary perspective with regards to ASAT warnings in open source projects.
\end{itemize}

The subjects of our case study are Java open source projects under the umbrella of the Apache Software Foundation.
We observe ASAT warning trends via PMD\footnote{https://pmd.github.io/, last accessed: 2019-04-11} and defects via the Issue Tracking System (ITS) of the respective projects under study.
In accordance with evidence based software engineering as introduced by~\cite{Kitchenham2004} we provide our data and analysis for researchers and practitioners regarding the evolution of warnings and the impact of PMD on software quality.

\rev{The main findings of our study are the following.
\begin{itemize}
    \item While the number of ASAT warnings is continuously increasing, the density of warnings per line of code is decreasing. 
    \item Most ASAT warning changes are related to style changes.
    \item The presence of PMD in the build file coincides with reduced defect density.
\end{itemize}}

The remainder of this work is structured as follows.
In Section~\ref{sec:related_work}, we discuss prior work related to this study. 
After that, in Section~\ref{sec:background}, we present a short overview of static analysis in software development and discuss challenges in mining software repository data. 
In Section~\ref{sec:case_study}, we define our research questions, describe the selection criteria for our study subjects and explain our methodology in detail. 
In Section~\ref{sec:results}, we present the results. In Section~\ref{sec:discussion}, we discuss the results and relate them to current research.
In Section~\ref{sec:threats}, we evaluate the threats to validity to our study. Section~\ref{sec:summary} provides a short conclusion and provides an outlook on future work based on the data and methods described this article.

\section{Related work}
\label{sec:related_work}
In this section, we present the related work on empirical studies of ASATs and put them into relation to our work.
\cite{state_static} investigated the usage of ASATs in open source projects. They focused on the prevalence of the usage of ASATs for different programming languages, how they are configured, and how the configuration evolves.
In our work we are also investigating the evolution of the configuration. In contrast to Beller et al., we also run an ASAT on our study subjects for each commit. This enables us to analyze when ASAT warnings are resolved or introduced and the kind and number of warnings. We are using the projects buildfiles to extract whether PMD or other ASATs are used at the time of the commit and if custom rulesets were deployed. Thus, we expand on the previous work by not only investigating the changes in the configuration but also if the ASAT was used to remove any warnings at all. The drawback of this detailed view on ASAT warnings is that we have to narrow the focus on one programming language and one ASAT.

\cite{warnings_static} utilized commit histories of ASAT warnings. They investigated the possibility of leveraging the removal times to prioritize the warnings. 
Instead of prioritizing ASAT warnings for removal, we are interested in removals on a global scale, by taking a longer history of the projects into account to get a broader view on the evolution of the projects under study with regard to ASAT warnings.

\cite{tse_findbugs} performed a large scale study using FindBugs\footnote{http://findbugs.sourceforge.net/, last accessed: 2019-04-10} via SonarQube\footnote{https://www.sonarqube.org/, last accessed: 2019-04-10} where they investigated ASAT warnings over time. They created an approach to identify fix-patterns that are then applied to unfixed warnings. Similar to our own work, Liu et al. have run an ASAT on the project source code retroactively. In comparison to Liu et al., we include the build system and custom rules in our analysis. Thus, we can be sure that when we investigate removal of warnings that the developers could have seen the warnings. Instead of FindBugs via SonarQube, we focus on PMD which reports a different set of warnings due to PMDs usage of source code instead of byte code. 

\cite{debt_apache} also utilized SonarQube to detect ASAT warnings and their removal. They focused on the technical debt metaphor~\citep{debt_metaphor} and the resolution time that SonarQube assigns to each detected ASAT warning. The authors took snapshots of their projects every two weeks to run the ASAT and store the warnings. In our study, we are not concerned with technical debt. Instead, we want to give a bigger, longitudinal overview over the evolution of the project regarding ASAT warnings. Instead of using snapshots, we ran PMD retroactively on every commit to extract data, although due to run time constraints, this results in a smaller number of projects in our study. Nevertheless, due to utilizing PMD our data covers a longer period of time.

\cite{marcelline} take a closer look at developer usage of ASATs through SonarQube.
They investigated the time to fix for different types of issues with a focus on active developer engagement to specifically solve the reported ASAT warnings.
In our study, we are not only concerned with resolution times. Instead, we are primarily interested in general trends regarding ASATs to infer information about the evolution of software quality in our candidate projects.

\cite{static_workshop} utilized data collected by~\cite{zimmermann1} and correlated source code quality metrics and defects with warnings found by different static analysis tools. They used three releases of eclipse and presented correlations for different size, complexity and object oriented source code metrics. In contrast to Plösch et al. we are not concerned only with releases, we collected static analysis warnings for every commit of our candidate projects. In addition, we consider multiple projects instead of one. Although we are only able to provide data for one static analysis tool, we are able to provide more detailed defect information and on a larger scale. This should also cover effects of readability and maintainability changes due to ASAT usage.

\rev{\cite{Querel2018} builds additional static analysis upon CommitGuru~\citep{rosen2015}. Initial results show that the additional information that static analysis warnings provide can improve statistical bug prediction models.
In our study, we investigate the evolution of ASAT warnings. Our own investigation into the impact of static analysis warnings on quality complements the initial results by~\cite{Querel2018}.}

\rev{Static analysis software is often used in a dedicated security context.
\cite{Penta2009} analyze security related ASAT warnings for three open source projects along their history. The authors performed an empirical study using three open source projects and three ASATs.}
\rev{\cite{Aloraini2019} also analyze security related ASAT warnings. The authors collect two snapshots, one at 2012 and one at 2017 for 116 open source projects.
Both works come to the conclusion that the warning density of the security related warnings stays constant throughout their analysis time span.
In contrast to our study both focused exclusively on security related ASATs.}

\section{Background}
\label{sec:background}
In this section, we introduce important topics regarding this study. First, we give a short description of the challenges regarding mining software repository histories and how they apply to this study. Then, we briefly discuss static analysis tools for Java and our choice of ASAT as well as software quality evaluation.

\subsection{Mining Software Repository Histories}
Working with old software revisions has its challenges. For projects which use the Java programming language some of these are:

\begin{itemize}
    \item The build system may have been switched completely, e.g., from Ant to Maven.
    \item The project has no pinned version for the libraries it needs to be built successfully. This means, it may be impossible to build an older version because of incompatibilities with required libraries or missing versions of libraries~\citep{oldversion_nocompile}.
    \item The main source directory may have been moved, e.g., from src/java to src/main/java as is the case for most Java projects with a longer history.
\end{itemize}

In this study, we follow two different paths of inquiry, the first is only concerned with general trends regarding ASAT warnings. Hence, we do not need to consider build systems and libraries. However, even in that simplest approach we \del{have to consider}\rev{ignore} test code as we only want to inspect production code.
As no direct information via the build system is available we utilize regular expression\rev{s} to exclude non-production code.

The second path of inquiry provides a more detailed view and also takes the build system into account. This is necessary as we extract ASAT usage via the build system configuration files.
We therefore restrict the build system to Apache Maven as it is used by the majority of our candidate projects and allows extraction of this information. Including build information provides us with the ability to restrict the production code via the source directories specified in the configuration.
The restriction to production code not only excludes test code but also additional tooling and examples. Apache Maven allows a tree like build configuration, i.e., a root configuration shared by the project and all its modules.
As the build configuration can also contain custom rules and ASAT configurations, we have to consider all parts of the configuration tree. The root of the tree, usually parent POMs, can be included via Maven central and the leaves, usually modules that are part of the project, can be included via the filesystem.

In addition to the build system, we restrict ourselves to an ASAT that does not need compiled source code because of preliminary tests which found similar problems as~\cite{oldversion_nocompile}. Tufano et al. found that 38\% of commits in their data could not be compiled anymore.
Nevertheless, even without the need to compile to bytecode, we are also experiencing some of the problems Tufano et al. found. As we want to extract custom rule definitions for PMD we need to consider build configuration files that may not exist anymore, e.g., missing parent POMs.
To mitigate this problem, we manually rename some artifacts so that they can be found on maven central and incorporated into our extraction process, usually this only consists of removing -SNAPSHOT from the package name but in \del{some}\rev{3} cases we need to change the name of the package, e.g., from commons to commons-parent.

\rev{The extraction first tries to build the effective POM via Maven, i.e., including every module and configuration as well as explicit default values.
If this fails it changes the pom.xml by removing the -SNAPSHOT, or, if the combination from group, artifact and version is in our rename list, it performs the artifact rename. After that the effective POM is built again.
In case that the error persists it is logged and the existing state of the custom rules is not changed. The remaining errors consist of Maven configuration mismatches, in most cases a module references a parent with a wrong version because the parent pom.xml has been upgraded but the modules still references the old version.}

\subsection{Static analysis tools for Java}
\label{sec:static_analysis_tools}
\cite{state_static} noted, that most static analysis tools are in use for languages which are not compiled, because the compilation process includes certain static checks. Nevertheless, even Java and also C have some static analysis tools that can be utilized by practitioners to warn about potential problems in the source code.

As we are focusing on Java there are a few well known, open source static analysis tools for Java. One of the most prevalent is Checkstyle\footnote{http://checkstyle.sourceforge.net/, last accessed: 2019-04-10} which works directly on the source code and is mostly concerned with checking the code against certain predefined coding style guidelines.
Another one is FindBugs which works on compiled Java bytecode to find bugs and common coding mistakes, e.g., a clone() method that may return null.
SonarQube, a cloud based tool, has the ability to use the already described  static analysis tools and also defines its own rules, e.g., cognitive complexity for a method is too high. It relates the ASAT warnings to a resolution time via a formula depending on the warning and programming language.

In this work we are focusing on PMD which works on the Java source code and finds coding style problems, e.g., an if without braces, but also common coding mistakes, e.g., comparison of two String objects using ``=='' instead of the equals() method. 
PMD provides a broad set of rules from a wide range of categories. 
Moreover, PMD is available since 2002 and therefore has been in use for a long time. This results in more data for our analysis and in a mature ASAT for us to use.
The detailed \del{detailed} documentation and changelog allow us to keep track of which ASAT warnings were available at a certain point in time.

\subsection{Software quality evaluation}
Software quality is notoriously hard to measure~\citep{Kitchenham1996}. Since the beginning of investigating software quality it seemed clear that software quality consists of a combination of factors. The first models for software quality introduced by~\cite{boehm,mccall} also mirror this combination of quality factors.
Multiple quality factors are still in use throughout the subsequent ISO standards 9126 and 25010 and later quality models, e.g.,~\cite{quamoco,columbus}. \rev{\cite{fenton} as well as the ISO standards discern between internal and external quality.
Internal quality factors concern the source code, e.g., cyclomatic complexity~\citep{mccabe} or the process, e.g., the developers. External quality factors are on the customer facing side, e.g., defects, efficiency. Internal quality factors influence external quality factors, the problem is to evaluate which internal factors influence which external factors in which way.}

Let us assume that software quality is a combination of multiple factors, e.g., maintainability or efficiency.
If we want to automatically evaluate software quality, we need to find the concrete measurements that capture the corresponding factor. We then also need to know how to combine the measurements or factors together for the best approximation of software quality.
Instead of using metric measurements as approximations, we can instead use ASATs based on their rules. Some ASATs not only include warnings about possible defects but also directly maintainability related warnings, e.g., default should always come last in a switch, exception handling code should not be empty or class names in Java should be in CamelCase. The rules that trigger the ASAT warnings are based on real world experiences and best practices of the developers, therefore, we expect that they are important in an overall evaluation of software quality.
Although this means that any ASAT considered for general software quality evaluation should support a broad set of rules.

If we consider the ASATs introduced in the previous Section~\ref{sec:static_analysis_tools} we find that PMD and SonarQube fit that definition best.
While FindBugs and Checkstyle are both very established software products they fit different profiles. FindBugs focuses on possible defects and Checkstyle focuses on validating style rule conformance.
While SonarQube would be a good fit, it does not exist for as long as PMD, FindBugs and Checkstyle. This limits the ability to observe actual usage of the ASAT in historical data.
PMD on the other hand has both a long history of use and a broad set of rules.
Therefore, we assume that PMD is a good approximation of internal software quality.

\rev{As an approximation for external software quality we utilize defect density~\citep{fenton}, i.e., the number of defects in relation to the size of the project. With both of these approximations, we can investigate internal and external software quality over the history of our study subjects.}

\section{Case study design}
\label{sec:case_study}
The \emph{goal} of the case study is to investigate evolution of ASAT warnings and to examine the impact of PMD in the short and long term on ASAT warning trends as well as its impact on external software quality via defect density.
In this Section, we formulate the research questions we aim to answer, explain the selection of subjects of the case study, and describe the methodology for the data collection, and the analysis procedures.

\subsection{Research questions}
\label{sec:research_questions}
To structure our investigations, we define the following research questions which we separate into two main questions.
The first main research question is only concerned with evolution of ASAT warnings over the full lifetime of the project: \textbf{How are ASAT warnings evolving over time?} (\emph{RQ1}).
We divide this research question into two sub-questions:
\begin{itemize}
    \item \emph{RQ1.1}: Is the number of ASAT warnings generally declining over time?
    \item \emph{RQ1.2}: Which warning types have declined or increased the most over time?
\end{itemize}
Investigating these questions should shed some light on the general evolution of our study subjects regarding ASAT warnings. More specifically, we want to answer the question if ``code gets better over time'' with regard to ASAT warnings and also if there are differences between the different types of ASAT warnings.
Differences between types of ASAT warnings may point to changing Java programming practices or changes in perceived importance, e.g., more camel case name violations for class names at the beginning of 2001 than at the end of 2017.
The trend of resolved warnings by type should indicate which warning types are perceived as the most important by the developers that are active in our study subjects.
\rev{ASAT warnings is a generic term, we specifically investigate ASAT warnings generated by PMD.}

The second research question is focused on the impact of ASAT usage on the warning trends and on external software quality: \textbf{What is the impact of using PMD?} (\emph{RQ2}).
We divide this research question into five sub-questions:
\begin{itemize}
    \item \emph{RQ2.1}: What is the short term impact of PMD on the number of ASAT warnings?
    \item \emph{RQ2.2}: What is the long term impact of PMD on the number of ASAT warnings?
    \item \emph{RQ2.3}: Does the active usage of custom rules for PMD correlate with higher ASAT warning removal?
    \item \emph{RQ2.4}: Is there a difference in ASAT warning removal trends whether PMD is included in the build process or not?
    \item \emph{RQ2.5}: Is there a difference in defect density whether PMD is included in the build process or not?
\end{itemize}
Our second set of research questions focuses on the ASAT usage according to the buildfile of the study subjects. Therefore, we focus only on the project development lifetimes where we can determine that an ASAT is used as part of the build process. 
Moreover, we consider only source directories configured in the build system. This allows us to exclude examples and tooling. We are also taking time and available rules for PMD into account, e.g., which rules are active in the configuration file and which were available at the time of the commit.
This enables us to analyze the impact only for the rules that the developers were able to see and therefore address consciously.
Moreover, we investigate the impact of PMD on external software quality via defect density by including information from the issue tracking system of our study subjects.

\subsection{Subject selection}
\label{sec:project_selection}
\begin{table}
    \centering
    \begin{tabular}{l|l}
        Criteria & Criteria category\\
        \hline
        \rev{at least one year issue tracker activity (from 1.1.2018 backwards)} & Project maturity\\
        at least one year development (from 1.1.2018 backwards) & Project maturity\\
        at least 1000 Commits & Project maturity\\
        no incubator & Project maturity\\
        commit activity since 1.1.2018 & Up-to-dateness\\
        issue activity since 1.1.2018 & Up-to-dateness\\
        at least 100 Files & Size\\
        uses Maven & Scope\\
        no Android & Scope\\
    \end{tabular}
    \caption{Project selection criteria}
    \label{tbl:selection_criteria}
\end{table}
Our study subjects are part of a convenience sample of open source projects under the leadership of the Apache Software Foundation\footnote{https://www.apache.org, last accessed: 2019-11-10} but nevertheless we applied some restrictions on our selection of study subjects.
\rev{The base list of projects consists of every Java project of the Apache Software Foundation. We then apply the restrictions and start mining the remaining projects. The final list of study subjects is a sample of the projects that pass the criteria. The complete data for all cannot be used due to the computational effort required to calculate the ASAT warnings for all commits.}

We focus on Java projects but exclude Android projects because of the different structure of the source code of the applications. We also restrict the build \rev{system} to Maven as we utilize the buildfiles to extract the source directory and ASAT configurations for \emph{RQ2}\del{ and}\rev{. Moreover,} Maven provides tooling necessary to combine multiple buildfiles of all sources per project.

We only include active, recent projects that are not currently in incubator status within the Apache Software Foundation, i.e., fully integrated into the Apache Software Foundation. All projects are actively using an issue tracking system as part of their development process.
Our study subjects consist of libraries and applications with a variety of domains, e.g., math libraries, pdf processing, http processing, machine learning, \rev{a} web application framework, and a wiki system. Moreover, our study subjects contain a diverse set of project sizes. The size ranges from small projects such as commons-rdf to lager projects such as Jena and Archiva.

The rest of our selection criteria are focused around project size, infrastructure, project maturity, and up-to-dateness of the project.
All applied criteria are given in Table~\ref{tbl:selection_criteria}.

\subsection{Methodology RQ1}
\label{sec:approach}
\begin{figure*}
    \centering
    \begin{tikzpicture}[node distance=0.7cm]

        \node at (-3.1, -2.05) {Select commit path};
        \node at (0, -2) {Metric extraction};
        \node[align=center] at (2.9, -2) {Calculate\\warning density};
        \node at (6, -2.05) {Fit linear regression};

        \node[single arrow, draw=none, fill=blue!20] at (-1.45, 0) {~~~};

        \node at (0, 0) {\includegraphics[width=2cm]{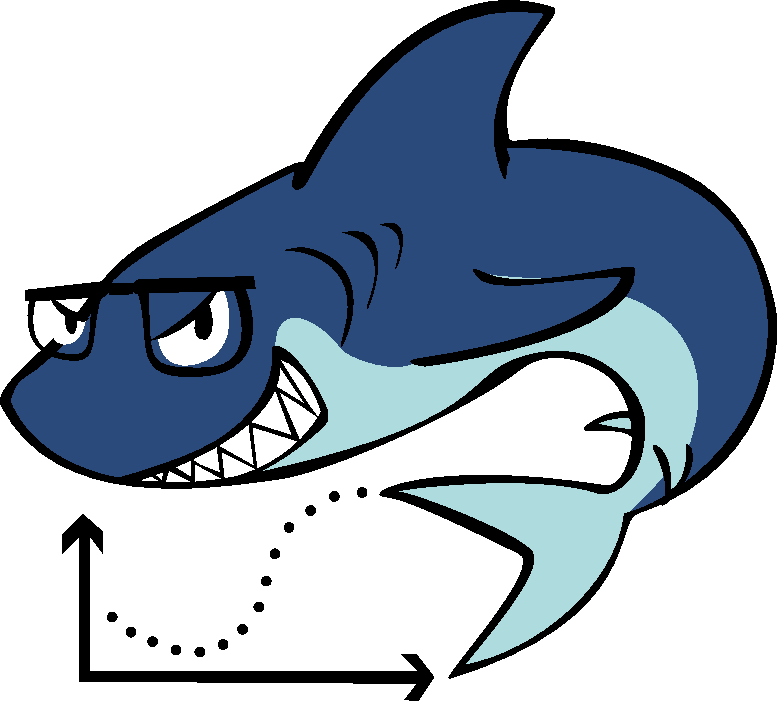}};

        \node[commit] at (-3.5, 0) (c1) {1};
        \node[commit, left of = c1] (c0) {0};
        \node[commit, above of = c1] (c2) {2};
        \node[commit, right of = c2] (c4) {4};
        \node[commit, right of = c1] (c3) {3};
        \node[commit, right of = c3] (c5) {5};

        \draw[->] (c0) -- (c1);
        \draw[->] (c1) -- (c2);
        \draw[->] (c2) -- (c4);
        \draw[->] (c1) -- (c3);
        \draw[->] (c3) -- (c5);

        \node[commit] at (-4.2, -1) (c10) {0};
        \node[commit, right of = c10] (c11) {1};
        \node[commit, right of = c11] (c13) {3};
        \node[commit, right of = c13] (c15) {5};

        \draw[->] (c10) -- (c11);
        \draw[->] (c11) -- (c13);
        \draw[->] (c13) -- (c15);

        \draw[decorate, decoration={brace,amplitude=10,mirror}] (-4.5,-0.3) -- (-1.8,-0.3);

        \node[single arrow, draw=none, fill=blue!20] at (1.4, 0) {~~~};

        \node at (2.8, 0) {$\text{R}=\frac{\sum\text{warnings}}{\sum\text{kLLoC}}$};

        \node[single arrow, draw=none, fill=blue!20] at (4.1, 0) {~~~};

        \node at (5.9, 0) {\includegraphics[width=3cm]{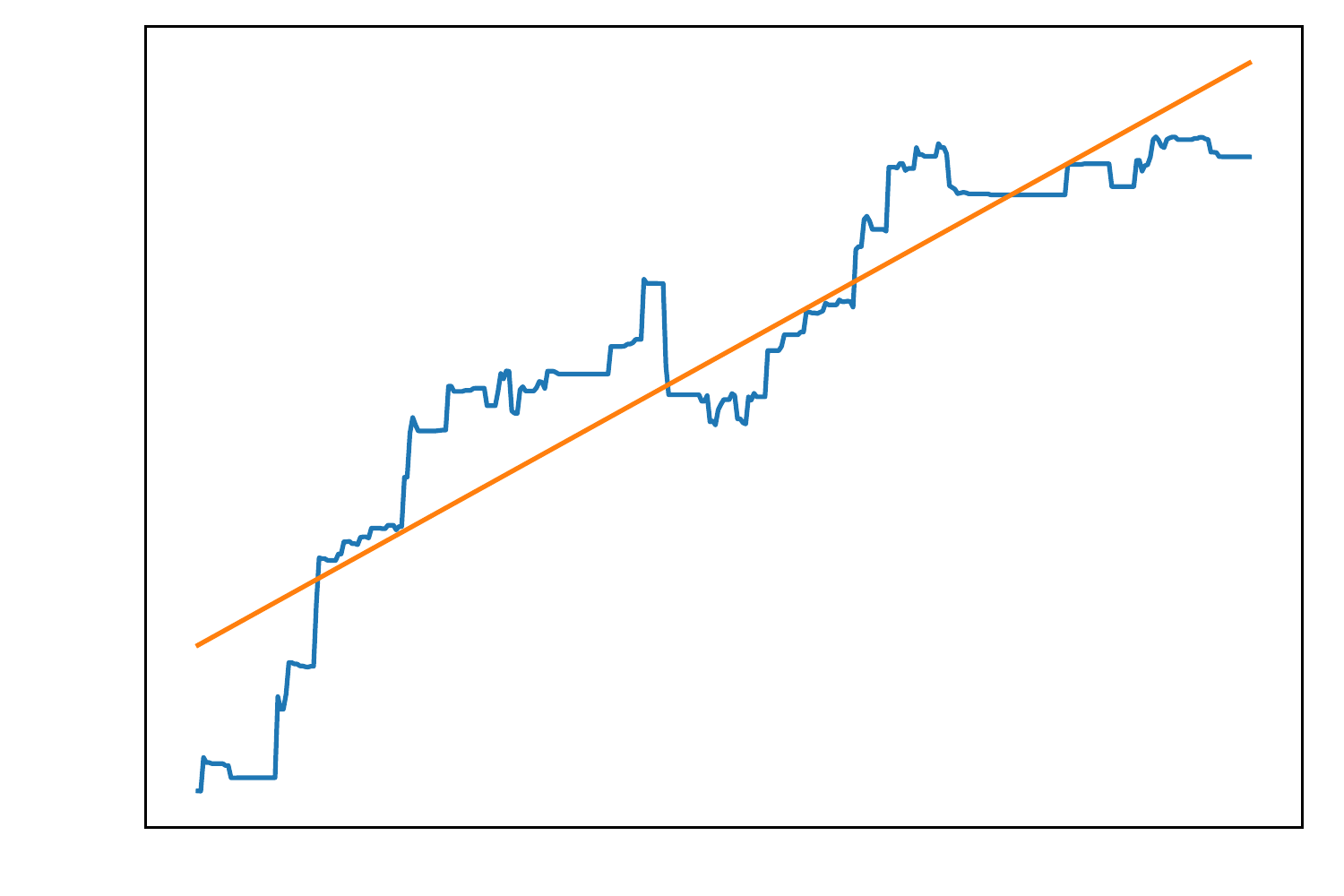}};

    \end{tikzpicture}
    \caption{Methodology RQ1}
    \label{fig:approach}
\end{figure*}
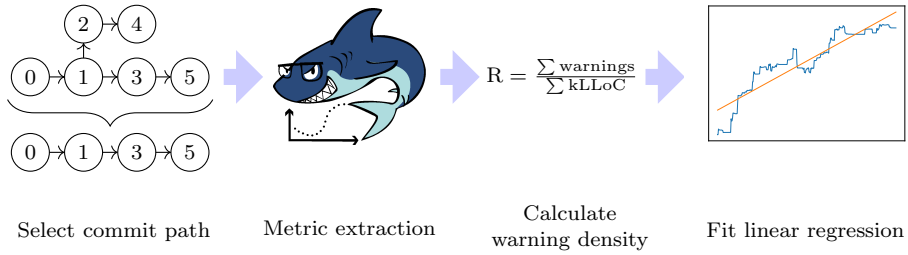
In this section, we explain our approach to extract the required data and to calculate the required metrics to answer our research questions.
An overview of the approach for data extraction is given in Figure~\ref{fig:approach}.

\subsubsection{Select commit path}
To select the commits we are interested in, we build a Directed Acyclic Graph (DAG) from all commits in the repository and their parent-child relationships. After the graph construction, we extract a single path of commits for the project. This is depicted in the first part of Figure~\ref{fig:approach}. Commits are denoted as circles with a number referring to their order of introduction into the codebase. We extract a single path from the latest master branch commit to the oldest reachable orphan commit. We need to select a path this way because we can not just select the master branch as the information on which branch a commit is created is not stored in Git~\citep{perils}. Moreover, we select a single path because if work is done in parallel on two or more branches of the project and we order the commits by date we get jumps in the data as we would have a sequence of commits that represent different states of the codebase at the same time. 

The latest master branch commit is extracted via the ``origin/head'' reference of Git which points to the default branch of the repository. The default branch is usually named master, although in some Apache projects the default branch is called trunk as the projects were converted or are mirrored from Subversion.
Orphan commits do not have parents. This is usually the initial commit of the repository. It can also happen that a repository has multiple orphan commits, which also can be merged back into the current development branch.
By choosing the oldest orphan commit, we extract the first initial commit. Then, we use the graph representation to find the shortest path between these two commits via Dijkstra's shortest path first algorithm~\citep{shortestpath}.
The end result of this step is the shortest path between the oldest orphan commit and the newest commit on the default branch of the project.

\subsubsection{Metric extraction}
\begin{table}
    \begin{tabular}{l|p{9cm}}
        File type & Regular expression\\
        \hline
        Test & \verb!(^|\/)(test|tests|test_long_running|testing|legacy-tests! \verb!|testdata|test-framework|derbyTesting|unitTests|java\/stubs! \verb!|test-lib|src\/it|src-lib-test|src-test|tests-src|test-cactus! \verb!|test-data|test-deprecated|src_unitTests|test-tools|! \verb!gateway-test-release-utils|gateway-test-ldap|nifi-mock)\/!\\
        Documentation & \verb!(^|\/)(doc|docs|example|examples|sample|samples|demo|tutorial! \verb!|helloworld|userguide|showcase|SafeDemo)\/!\\
        Other & \verb!(^|\/)(_site|auxiliary-builds|gen-java|external! \verb!|nifi-external)\/!\\
    \end{tabular}
    \caption{\rev{Regular expressions for excluding non-production code.}}
    \label{tbl:regex}
\end{table}
The second step in Figure~\ref{fig:approach} depicts the extraction of ASAT warnings and Software metrics.
For both we are using OpenStaticAnalzer\footnote{https://github.com/sed-inf-u-szeged/OpenStaticAnalyzer/, last accessed: 2019-11-10} as part of a plugin\footnote{https://www.github.com/smartshark/mecoshark/, last accessed: 2019-11-10} for the SmartSHARK infrastructure~\citep{smartshark}.
SmartSHARK in conjunction with a HPC-Cluster provided us with the means to extract this information for each file in each commit of our candidate projects.
\rev{OpenStaticAnaylzer is an open sourced version of the commercial tool SourceMeter~\citep{sourcemeter} which has been used in multiple studies, e.g., \cite{code_ownership,bulk_fixing,Ferenc2014} and, more recently \cite{Ferenc2020}.
It works by constructing an Abstract Semantic Graph (ASG) from the source code which is then used to calculate static source code metrics.
As it is included in SmartSHARK we perform a validation step after each mining step which verifies if the metrics are collected for each source code file.}
In addition to the size, complexity and coupling metrics OpenStaticAnalyzer also provides us with ASAT warnings by PMD. OpenStaticAnalyzer applies 193 rules from which the warnings are generated including line number, type and severity rating. The source code metrics are provided at package, file, class, method and attribute level. The resulting data from the mining step includes ASAT warnings from PMD and source code metrics for each file in each commit of our candidate projects.
As we primarily want to investigate program code we exclude \rev{non-production code} by path. \rev{We use the regular expression shown in Table~\ref{tbl:regex} to filter non-production code. 
The regular expressions were created based on manual inspection of the directory structure of the project we use in our study.}

Furthermore, we only compare full years of continuous development in our analysis, thus we remove incomplete years: we remove the first year and everything after 31.12.2017 because we started collecting the data in 2018.
This ensures that we only have trends over the complete development history of the project but also for each single year of development which provides a more detailed view in addition to a full view of the projects lifetime.

\subsubsection{Calculate warning density}
\label{sec:warning_density}
The absolute number of ASAT warnings is \del{strongly} correlated with the amount of source code in the project. Increasing the code size \del{usually}\rev{seems to} increase\del{s} the number of warnings. Even in projects using PMD, this is expected as we also study warnings which the developers could not have seen before. Either because the ASAT did not support them at the time of the commit or the rules that trigger the warnings are not active. Most of the biggest additions and removals of ASAT warnings are due to the addition and removal of files in the repository. The measure for size of the source code we are using is Logical Lines of Code (LLoC) in steps of one thousand (kLLoC). By using LLoC instead of just Lines of Code (LOC) we discard blank lines and comments. LLoC provides a more realistic estimation of the project size.

Table~\ref{tbl:correlation} shows the correlation between the sum of ASAT warnings and the sum of kLLoC per commit in all commits available in our data.
We are using two non-parametric correlation metrics, Kendall's $\tau$~\citep{kendall}, which uses concordance of pairs, i.e., if $x_i > x_j$ and $y_i > y_j$ and Spearman's $\rho$~\citep{spearman} which uses a rank transformation to measure the monotonicity between two sets of values instead of concordance of pairs of observations.

\begin{table}
\centering
\begin{tabular}{l|c|c}
    Method & Value & P-value\\
    \hline
    Kendall's $\tau$ & \del{0.55976}\rev{0.57509} & 0.0\\
	Spearman's $\rho$ & \del{0.69857}\rev{0.71654} & 0.0\\
\end{tabular}
\caption{Correlation between the number of ASAT warnings and kLLoC}
\label{tbl:correlation}
\end{table}

We can see in Table~\ref{tbl:correlation} that there is a \del{medium to strong} positive correlation between kLLoC and the number of ASAT warnings, i.e., as kLLoC increases so does the number of ASAT warnings.
As we want to analyze ASAT warning trends with minimum interference of functionality being added or deleted we decided to use warning density instead of the absolute number of ASAT warnings. Warning density is the ratio of the ASAT warnings and product size.

\begin{equation}
    \text{Warning density} = \frac{\text{Number of ASAT warnings}}{\text{Product size}}
\end{equation}
As product size we chose kLLoC, the warning density is calculated per commit.
The advantage of this measurement is that we still see when code with less warnings is added or removed. This also accounts for the effect of developers only scrutinizing new code being added as the new code would then contain less warnings than the existing and show up in our data as a declining trend of ASAT warnings.

Nevertheless, we keep the sum of all warnings for completeness which means we have two aggregations of warning data for our next step:

\begin{description}
    \item[\emph{S} (sum):] The sum of all ASAT warnings per commit which are also available on basis of warning type and severity rating.
    \item[\emph{R} (warning density):] The ratio meaning the warning density per commit which is also available on a basis of warning type and severity rating.
\end{description}

\subsubsection{Fit linear regression}
Fitting a regression line results in a trend line that we can use to determine if ASAT warnings are generally increasing or decreasing in a more appropriate way as just using a delta between the last and first data points.
This method, while still being simple and comprehensible, utilizes all available information, e.g., if the project contained high numbers of warnings for most of its lifetime and only at the end of the extracted data resolved most of them.
We fit multiple linear regression lines to our data:
\begin{itemize}
    \item all years per project, for a long-term trend,
    \item per year per project, for a short-term trend.
\end{itemize}
Moreover, we additionally fit regression lines for each group of filtered ASAT warnings we introduce in Section~\ref{sec:filter_warnings} for our second main research question.
The linear regression lines provide broad overall trends and specific trends for the ASAT warnings to answer our research questions.

After the fitting of the regression lines, we utilize the coefficient of the linear regressions as the slope. As we have only one variable, this is the same as calculating the slope for each line by applying the point slope formula (Equation~\ref{eq:slope}) where $y$ are the values of the fitted regression line and $x$ is the day of the commit.
\begin{equation} \label{eq:slope}
    \text{slope} = \frac{y_n - y_1}{x_m - x_1}
\end{equation}

The slope provides us with a single number representing the trend which we use for further analyses. Moreover, this enables the merging of results for projects with different lifetimes in order to create a global overview of a trend.

In order to restrict the calculated trends to meaningful values we use an F-Score which is calculated via a correlation between our regression line and the measured value.
First, we calculate the correlation:

\begin{equation}
    \text{corr} = \frac{(X_i - \overline{X}) \cdot (y_j - \overline{y})}{\sigma(X) \cdot \sigma(y)}
\end{equation}

Where $X_i$ is the $i$-th day of our commits, $y_j$ is the $j$\hbox{-}th value of the regression line and $\overline{X}, \overline{y}$ is the mean of the number of days of commits and mean of the regression values respectively. $\sigma(X), \sigma(y)$ denotes the standard deviation of $X$ and $y$.
The correlation is then converted to an F-Score and a p-value.
\begin{equation}
    \text{F} = \frac{\text{corr}^2}{(1 - \text{corr}^2) \cdot (|y| - 2)}
\end{equation}

The p-value conversion is achieved via the survival function of the F-distribution.

To restrict noise introduced by bad regression fits for trends, we include only slope values in our analysis where the F-Score is above $1$ and the p-value for the F-Score is lower than $0.05$. As the F-Score describes a relationship between the regression values and the time, we chose this performance metric instead of others related to linear regression such as $R^2$.

\subsection{Methodology RQ2}
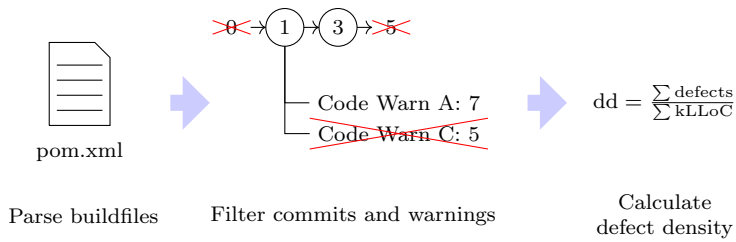
\begin{figure}
    \centering
    \begin{tikzpicture}[node distance=0.7cm]

        \node at (2.85, -2) {Parse buildfiles};
        \node at (6.4, -2) {Filter commits and warnings};
        \node[align=center] at (10.5, -2) {Calculate\\defect density};

        \draw (2.4,-0.8) -- (3.2,-0.8) -- (3.2,0.1) -- (3.0,0.3) -- (2.4,0.3) -- (2.4,-0.8);
        \draw (2.5,-0.6) -- (3.1, -0.6);
        \draw (2.5,-0.4) -- (3.1, -0.4);
        \draw (2.5,-0.2) -- (3.1, -0.2);
        \draw (2.5,0) -- (3.1, 0);

        \node[align=center] at (2.8, -1.15) (files) {pom.xml};

        \node[single arrow, draw=none, fill=blue!20] at (4.2, -0.5) {~~~};

        \node[commit, cross out, color=red, text=black] at (4.8, 0.5) (c40) {0};
        \node[commit, right of = c40] (c41) {1};
        \node[commit, right of = c41] (c43) {3};
        \node[commit, right of = c43, cross out, color=red, text=black] (c45) {5};

        \draw[->] (c40) -- (c41);
        \draw[->] (c41) -- (c43);
        \draw[->] (c43) -- (c45);

        \node at (7, -0.5) (lloc) {Code Warn A: 7};
        \node[draw, below = 0cm of lloc, cross out, color = red, text = black] (pmd) {Code Warn C: 5};

        \draw (c41) |- (lloc);
        \draw (c41) |- (pmd);

        \node[single arrow, draw=none, fill=blue!20] at (8.9, -0.5) {~~~};

        \node at (10.5, -0.5) {$\text{dd}=\frac{\sum\text{defects}}{\sum\text{kLLoC}}$};

    \end{tikzpicture}
    \caption{Methodology extension for RQ2}
    \label{fig:approach2}
\end{figure}

To answer our second research question, we need to include knowledge about the inclusion of ASATs in the build process of the projects. As previously mentioned, we focus on Maven as the build system. To extract the additional information, we extend our approach shown in Figure~\ref{fig:approach} with the additional steps depicted in Figure~\ref{fig:approach2}. In a nutshell, we filter out commits where Maven was not used, create new sets of rules depending on custom rules included in the available Maven build configuration, and include defect density as external software quality measure.

\subsubsection{Parse buildfiles}
We traverse the path of commits previously selected to determine where Maven was introduced to the project and all commits where its configuration file was changed.
Maven projects can contain multiple buildfiles, modules and configuration residing in parent buildfiles. In order to account for these features, we utilize a Maven feature that combines all this information including fetching the parent buildfiles from the Maven repository.
For each commit where one or multiple Maven files were changed in the target repository, we execute Maven to automatically resolve potential project modules defined in the main Maven configuration file, potential parent configurations, and set all settings explicitly taking default values and overrides into account.
We also extract source and test directories from the configuration which allows us to restrict our analysis to program source code and to exclude tests that reside in non-standard directories.
This information is further refined to extract which ASATs are currently active, i.e., we detect if PMD, Checkstyle, and FindBugs are configured.
If PMD is configured, we extract the configuration including all additional custom configuration files.

Custom configurations for PMD can consist of multiple files with rules and categories of rules. We parse every custom ruleset file and extract rule categories and single rules. 
Single rules are used as is, whereas the rule categories are expanded to the single rules they contain according to the current PMD documentation on all rulesets\footnote{https://pmd.github.io/latest/pmd\_rules\_java.html\#additional-rulesets}.
This ensures that we have an accurate representation of the warnings that were actively presented to the developers.

\subsubsection{Filter commits and warnings}
\label{sec:filter_warnings}
We remove all commits where no Maven buildfile was present. This is true for commits where the build system is not Maven but, e.g., Ant or Gradle.
To make our comparisons viable, we restrict our data to Maven and remove commits until a Maven buildfile is introduced.
The project selection performed in the first step ensures that we only have projects where the Maven buildfile was present in the latest commit of our data. Thus, we do not have to remove commits due to the project switching its buildsystem from Maven to Gradle.
After that, we create subsets of warnings in our data by filtering certain warnings.

\begin{description}
    \item[\emph{t} (time-corrected)] The first subset consists of time corrected warnings. This includes only warnings that were available at the time of the commit where we collected the warning.
To be able to utilize this information, we included a mapping for each detected rule to the PMD version that introduced the warning together with the release date of that version.
Then we filter out the rules that could not have been reported because at the time of the commit the rule was not available in PMD yet.
This is only possible because of the very thorough documentation of rules from PMD and the detailed changelog that stretches back to the first version. 
We include this subset because of the length of the project histories considered. As some of our data goes back to 2007\footnote{Earliest date for which a Maven buildfile exists in our data.}, we need to take the ASAT warnings into account that were possible to gain from PMD at that point in time.

    \item[\emph{d} (default):] The second subset represents the default configuration of the maven-pmd-plugin.
The default rules are taken from the most recent configuration\footnote{https://maven.apache.org/plugins/maven-pmd-plugin/examples/usingRuleSets.html, last accessed: 2019-03-18} and filtered to include only detectable rules. This results in a set of 45 rules.

    \item[\emph{e} (effective):] For the third subset we want to include as much detail as possible. To achieve this, we calculate the currently active ASAT rules for each commit.
These effective active rules take all custom rules and rule excludes into account. If no custom rules are defined we are using the default rules according to the documentation of the maven-pmd-plugin\footnotemark[11], same as for the subset \emph{d}.
This subset contains all information that a developer on the project under investigation can acquire by utilizing the buildfile.

    \item[\emph{o} (without overlapping):] The final subset removes rules overlapping with other ASATs used in the projects. This is achieved by filtering rules which overlap with rules supported by current versions of Checkstyle and FindBugs.
This subset enables us to increase the precision of our impact measurement for the effects of using PMD on ASAT warning trends. \rev{This avoids skewed results for study subjects which are in the non PMD group but utilize FindBugs or Checkstyle which contain rules that are also present in PMD. A complete list of overlapping rules can be found in the Appendix.}
\end{description}

All of these subsets of warnings can be combined to provide us with a set of rules for the analyses, e.g., the warning density of the default rules with time-correction or the warning density of the effective rules with time-correction without overlapping rules.

\subsubsection{Calculate defect density}
\label{sec:defect_density}
This step utilizes the SmartSHARK infrastructure which we already used for the metrics collection to incorporate information from the ITS into our data.
The extracted information is condensed to a metric per development year, the ``de facto standard measure of software quality''~\citep{fenton}, defect density, which is a ratio of the number of known defects and the size of the product.

\begin{equation}
    \text{Defect density} = \frac{\text{Number of known defects}}{\text{Product size}}
\end{equation}

In this study we use the mean kLLoC per year as the product size and the number of created bug reports per year as the number of known defects.
This proves us with a metric per year which we can then utilize to measure the impact of PMD usage on external software quality~\citep{fenton}.
As we have projects in our dataset which switched ITS and as we need full development years we discard the first year for which we have defects in our data.

\subsection{Analysis procedure}
In our case study, we investigate different questions which require a different analysis procedures.
For \emph{RQ1.1}, we aggregate the plain sum of ASAT warnings per commit over the projects development and the warning density as defined in Section~\ref{sec:warning_density}.
Then, we fit regression lines and calculate the F-Score as well as the slope of the regression to get the general trend.

In the case of \emph{RQ1.2} we do the same, but for completeness we additionally calculate the delta of the last and the first commit of the data as well as the number of remaining warnings per kLLoC.
\rev{In order to aggregate data of all projects we calculate the mean and median of the data.}

The short and long term impact of PMD on the number of ASAT warnings in \emph{RQ2.1} and \emph{RQ2.2} is measured via the number of projects for \emph{RQ2.1} and the \rev{median} of slopes of the trend line for all years following PMD introduction per project for \emph{RQ2.2}.
The slopes are calculated via the warning density but without overlapping rules from FindBugs and Checkstyle.

For \emph{RQ2.3}, we sum the number of rule changes per year for projects using PMD and correlate them via Kendall's~$\tau$ and Spearman's~$\rho$ to the warning density trends of the rules used (\emph{R}+\emph{e}+\emph{t}).

To answer the research questions \emph{RQ2.4} and \emph{RQ2.5}, we measure the difference between two samples. We first investigate the distribution of our data via the Shapiro-Wilk test~\citep{shapiro_wilk} for normality and Levene's test~\citep{levene} for variance homogeneity.
As these tests revealed that the data is non-normal with a homogeneous variance, we decided to use the Mann-Whitney-U test~\citep{mwu}. Although the Mann-Whitney-U test is a ranked test we still talk about differences in median for the sake of simplicity. 
In both research questions, we measure the difference between the years of PMD usage and the years where PMD was not used. Partial use in a year is excluded from the analysis.
We chose a significance level of $\alpha = 0.05$, after Bonferroni~\citep{bonferroni} correction for 24 statistical tests, we reject the $H_0$ hypothesis at $p < 0.002$.
The difference in median between both samples is not significant every time for $p < 0.002$. Therefore, only for the last comparison of \emph{RQ2.4} and for \emph{RQ2.5} we also calculate effect size and confidence interval.

To calculate the effect size of the Mann-Whitney-U test, we utilize the fact that for sample sizes $>$ 8 the U test statistic is approximately normally distributed~\citep{mwu}.
\rev{We first perform a $z$-standardization \citep{Kreyszig}.
We are assuming that our sample's mean and standard deviation are a good approximation of the populations mean and standard deviation. After we calculate $z$ we can calculate the effect size $r$.
A value of $r < 0.3$ is considered a small effect, $0.3 \leq r \leq 0.5$ is a medium effect and $0.5 < r$ is a strong effect~\citep{cohen}.
For the confidence interval we follow~\cite{Campbell} who use the $K$-th smallest to the $K$-th largest difference between two samples as the interval.
The confidence interval then consists of the $K$-th difference and the $max(n,m)-K$-th difference between both samples.}







\subsection{Replication kit}
All extracted data can be found online~\rev{\citep{replication_kit}}.
The code for creation of the tables and figures used in this paper as well as a dynamic view of warning density, LLoC and warning sum for each project is included.

\section{Case study results}
\label{sec:results}
In this section we present the results of our study. This section \rev{is} split into two parts, one for each of our main research questions.

\subsection{\emph{RQ1}: How are ASAT warnings evolving over time?}
\label{sec:rq1}

Our first research question considers ASAT warning evolution over the complete lifetime of each project. \rev{We do not consider PMD usage in the build process, custom rulesets or the availability of warnings in PMD at the time of the commit in this section.  We use all PMD rules that are available.} 
Table~\ref{tbl:trends_full2} shows the trend of ASAT warnings for every project and year as well as the approximate change per year over all years. Furthermore, the table includes the trend over the complete lifetime of the project with two base values: the sum of ASAT warnings \emph{S} and the warning density \emph{R}. The trends for single years are calculated based on warning density.
The arrows indicate the trend of the ASAT warnings. A downwards arrow indicate a positive trend, i.e., the warning density declines, an upwards arrow negative trend, i.e., the warning density increases. 
If our criteria for the regression fits are not met, i.e., the F-Score is below $1$ and the corresponding p-value is above $0.05$ a straight rightway arrow is used. 
Hence, the straight rightway arrow indicates that there was no significant change.

\begin{sidewaystable*}
\captionsetup{font=scriptsize}
\hspace{-1.5cm}
\tiny
\begin{tabular}{l|c|c|c|c|c|c|c|c|c|c|c|c|c|c|c|c|c|c|c|r}
Project & 2001 & 2002 & 2003 & 2004 & 2005 & 2006 & 2007 & 2008 & 2009 & 2010 & 2011 & 2012 & 2013 & 2014 & 2015 & 2016 & 2017 & \emph{S} & \emph{R} & \emph{R} p.a.\\
\hline
archiva &   &   &   &   &   & $\searrow$ & $\nearrow$ & $\searrow$ & $\searrow$ & $\nearrow$ & $\nearrow$ & $\searrow$ & $\nearrow$ & $\searrow$ & $\searrow$ & $\nearrow$ & $\nearrow$ & $\nearrow$ & $\nearrow$ & 2.0638\\
calcite &   &   &   &   &   &   &   &   &   &   &   &   & $\nearrow$ & $\nearrow$ & $\nearrow$ & $\rightarrow$ & $\searrow$ & $\nearrow$ & $\nearrow$ & 1.4294\\
cayenne &   &   &   &   &   &   &   & $\nearrow$ & $\searrow$ & $\searrow$ & $\searrow$ & $\nearrow$ & $\nearrow$ & $\rightarrow$ & $\rightarrow$ & $\nearrow$ & $\nearrow$ & $\nearrow$ & $\rightarrow$ & -\\
commons-bcel &   & $\searrow$ & $\searrow$ & $\searrow$ & $\searrow$ & $\searrow$ & $\rightarrow$ & $\searrow$ & $\nearrow$ & $\searrow$ & $\searrow$ & $\nearrow$ & $\rightarrow$ & $\searrow$ & $\rightarrow$ & $\searrow$ & $\searrow$ & $\searrow$ & $\searrow$ & -5.6761\\
commons-beanutils &   & $\searrow$ & $\nearrow$ & $\searrow$ & $\searrow$ & $\searrow$ & $\searrow$ & $\rightarrow$ & $\nearrow$ & $\nearrow$ & $\rightarrow$ & $\rightarrow$ & $\searrow$ & $\searrow$ & $\rightarrow$ & $\searrow$ & $\nearrow$ & $\nearrow$ & $\searrow$ & -2.4991\\
commons-codec &   &   &   & $\nearrow$ & $\searrow$ & $\rightarrow$ & $\searrow$ & $\rightarrow$ & $\searrow$ & $\rightarrow$ & $\nearrow$ & $\nearrow$ & $\searrow$ & $\searrow$ & $\searrow$ & $\searrow$ & $\searrow$ & $\nearrow$ & $\searrow$ & -1.1347\\
commons-collections &   & $\rightarrow$ & $\rightarrow$ & $\searrow$ & $\rightarrow$ & $\searrow$ & $\rightarrow$ & $\nearrow$ & $\searrow$ & $\searrow$ & $\searrow$ & $\searrow$ & $\searrow$ & $\rightarrow$ & $\nearrow$ & $\searrow$ & $\searrow$ & $\searrow$ & $\searrow$ & -2.2388\\
commons-compress &   &   &   & $\searrow$ & $\rightarrow$ & $\nearrow$ & $\rightarrow$ & $\nearrow$ & $\searrow$ & $\rightarrow$ & $\searrow$ & $\searrow$ & $\nearrow$ & $\rightarrow$ & $\rightarrow$ & $\searrow$ & $\searrow$ & $\nearrow$ & $\searrow$ & -0.8231\\
commons-configuration &   &   &   & $\nearrow$ & $\nearrow$ & $\searrow$ & $\nearrow$ & $\searrow$ & $\searrow$ & $\searrow$ & $\searrow$ & $\searrow$ & $\searrow$ & $\searrow$ & $\searrow$ & $\nearrow$ & $\nearrow$ & $\nearrow$ & $\searrow$ & -1.9519\\
commons-dbcp &   & $\searrow$ & $\searrow$ & $\searrow$ & $\nearrow$ & $\searrow$ & $\searrow$ & $\searrow$ & $\searrow$ & $\searrow$ & $\searrow$ & $\rightarrow$ & $\nearrow$ & $\searrow$ & $\rightarrow$ & $\rightarrow$ & $\searrow$ & $\nearrow$ & $\searrow$ & -5.9619\\
commons-digester &   & $\searrow$ & $\searrow$ & $\nearrow$ & $\searrow$ & $\searrow$ & $\searrow$ & $\rightarrow$ & $\rightarrow$ & $\searrow$ & $\searrow$ & $\searrow$ & $\searrow$ & $\rightarrow$ & $\rightarrow$ & $\rightarrow$ & $\rightarrow$ & $\rightarrow$ & $\searrow$ & -7.9792\\
commons-imaging &   &   &   &   &   &   &   & $\searrow$ & $\rightarrow$ & $\searrow$ & $\searrow$ & $\searrow$ & $\searrow$ & $\rightarrow$ & $\nearrow$ & $\nearrow$ & $\rightarrow$ & $\searrow$ & $\searrow$ & -15.7747\\
commons-io &   &   & $\searrow$ & $\nearrow$ & $\searrow$ & $\nearrow$ & $\nearrow$ & $\rightarrow$ & $\rightarrow$ & $\searrow$ & $\searrow$ & $\nearrow$ & $\searrow$ & $\searrow$ & $\searrow$ & $\nearrow$ & $\searrow$ & $\nearrow$ & $\searrow$ & -0.6867\\
commons-jcs &   &   & $\searrow$ & $\nearrow$ & $\searrow$ & $\searrow$ & $\nearrow$ & $\searrow$ & $\searrow$ &   & $\searrow$ & $\searrow$ & $\rightarrow$ & $\searrow$ & $\searrow$ & $\rightarrow$ & $\rightarrow$ & $\nearrow$ & $\searrow$ & -3.0907\\
commons-jexl &   &   & $\searrow$ & $\searrow$ & $\nearrow$ & $\searrow$ & $\rightarrow$ & $\searrow$ & $\searrow$ & $\nearrow$ & $\searrow$ & $\nearrow$ & $\rightarrow$ & $\rightarrow$ & $\searrow$ & $\rightarrow$ & $\searrow$ & $\searrow$ & $\searrow$ & -9.5675\\
commons-lang &   &   & $\searrow$ & $\searrow$ & $\searrow$ & $\searrow$ & $\searrow$ & $\nearrow$ & $\nearrow$ & $\searrow$ & $\searrow$ & $\searrow$ & $\searrow$ & $\searrow$ & $\nearrow$ & $\searrow$ & $\searrow$ & $\nearrow$ & $\searrow$ & -1.8343\\
commons-math &   &   &   & $\searrow$ & $\searrow$ & $\nearrow$ & $\searrow$ & $\searrow$ & $\searrow$ & $\searrow$ & $\searrow$ & $\searrow$ & $\searrow$ & $\searrow$ & $\searrow$ & $\rightarrow$ & $\nearrow$ & $\nearrow$ & $\searrow$ & -5.4131\\
commons-net &   &   & $\searrow$ & $\nearrow$ & $\searrow$ & $\searrow$ & $\rightarrow$ & $\searrow$ & $\searrow$ & $\searrow$ & $\searrow$ & $\searrow$ & $\searrow$ & $\searrow$ & $\searrow$ & $\searrow$ & $\searrow$ & $\searrow$ & $\searrow$ & -4.6747\\
commons-rdf &   &   &   &   &   &   &   &   &   &   &   &   &   &   & $\searrow$ & $\nearrow$ & $\nearrow$ & $\nearrow$ & $\nearrow$ & 9.0116\\
commons-scxml &   &   &   &   &   & $\nearrow$ & $\nearrow$ & $\searrow$ & $\searrow$ & $\rightarrow$ & $\searrow$ & $\rightarrow$ & $\searrow$ & $\searrow$ & $\searrow$ & $\rightarrow$ & $\nearrow$ & $\nearrow$ & $\searrow$ & -1.9494\\
commons-validator &   &   & $\searrow$ & $\nearrow$ & $\searrow$ & $\searrow$ & $\searrow$ & $\rightarrow$ & $\rightarrow$ & $\rightarrow$ & $\nearrow$ & $\searrow$ & $\searrow$ & $\searrow$ & $\nearrow$ & $\searrow$ & $\nearrow$ & $\nearrow$ & $\searrow$ & -3.7802\\
commons-vfs &   &   & $\searrow$ & $\nearrow$ & $\nearrow$ & $\nearrow$ & $\searrow$ & $\nearrow$ & $\searrow$ & $\searrow$ & $\nearrow$ & $\rightarrow$ & $\nearrow$ & $\searrow$ & $\searrow$ & $\searrow$ & $\nearrow$ & $\nearrow$ & $\searrow$ & -0.5570\\
eagle &   &   &   &   &   &   &   &   &   &   &   &   &   &   &   & $\searrow$ & $\searrow$ & $\nearrow$ & $\searrow$ & -22.7229\\
falcon &   &   &   &   &   &   &   &   &   &   &   & $\nearrow$ & $\searrow$ & $\searrow$ & $\searrow$ & $\nearrow$ & $\nearrow$ & $\nearrow$ & $\searrow$ & -8.9550\\
flume &   &   &   &   &   &   &   &   &   &   &   & $\nearrow$ & $\rightarrow$ & $\searrow$ & $\searrow$ & $\searrow$ & $\nearrow$ & $\nearrow$ & $\searrow$ & -0.1371\\
giraph &   &   &   &   &   &   &   &   &   & $\searrow$ &   & $\searrow$ & $\nearrow$ & $\searrow$ & $\searrow$ & $\searrow$ & $\searrow$ & $\nearrow$ & $\searrow$ & -7.8048\\
gora &   &   &   &   &   &   &   &   &   &   & $\searrow$ & $\nearrow$ & $\nearrow$ & $\searrow$ & $\searrow$ & $\searrow$ & $\searrow$ & $\nearrow$ & $\nearrow$ & 2.2756\\
helix &   &   &   &   &   &   &   &   &   &   &   & $\searrow$ & $\nearrow$ & $\searrow$ & $\searrow$ & $\searrow$ & $\searrow$ & $\nearrow$ & $\nearrow$ & 0.5769\\
httpcomponents-client &   &   &   &   &   & $\nearrow$ & $\searrow$ & $\nearrow$ & $\nearrow$ & $\nearrow$ & $\searrow$ & $\searrow$ & $\searrow$ & $\searrow$ & $\nearrow$ & $\rightarrow$ & $\nearrow$ & $\nearrow$ & $\searrow$ & -2.0253\\
httpcomponents-core &   &   &   &   &   & $\searrow$ & $\searrow$ & $\searrow$ & $\searrow$ & $\nearrow$ & $\nearrow$ & $\searrow$ & $\searrow$ & $\nearrow$ & $\searrow$ & $\searrow$ & $\searrow$ & $\nearrow$ & $\searrow$ & -1.1508\\
jena &   &   &   &   &   &   &   &   &   &   &   &   & $\searrow$ & $\searrow$ & $\nearrow$ & $\nearrow$ & $\nearrow$ & $\nearrow$ & $\searrow$ & -5.4301\\
jspwiki & $\nearrow$ & $\searrow$ & $\searrow$ & $\nearrow$ & $\searrow$ & $\searrow$ &   & $\searrow$ & $\searrow$ & $\nearrow$ & $\nearrow$ & $\searrow$ & $\searrow$ & $\nearrow$ & $\rightarrow$ & $\searrow$ & $\searrow$ & $\nearrow$ & $\searrow$ & -3.9080\\
knox &   &   &   &   &   &   &   &   &   &   &   &   & $\searrow$ & $\searrow$ & $\nearrow$ & $\searrow$ & $\searrow$ & $\nearrow$ & $\searrow$ & -2.9519\\
kylin &   &   &   &   &   &   &   &   &   &   &   &   &   &   & $\nearrow$ & $\nearrow$ & $\searrow$ & $\nearrow$ & $\searrow$ & -0.2880\\
lens &   &   &   &   &   &   &   &   &   &   &   &   &   & $\rightarrow$ & $\searrow$ & $\nearrow$ & $\searrow$ & $\nearrow$ & $\searrow$ & -8.4280\\
mahout &   &   &   &   &   &   &   &   & $\nearrow$ & $\searrow$ & $\searrow$ & $\searrow$ & $\nearrow$ & $\nearrow$ & $\nearrow$ & $\nearrow$ & $\nearrow$ & $\rightarrow$ & $\searrow$ & -4.0993\\
manifoldcf &   &   &   &   &   &   &   &   &   &   & $\searrow$ & $\rightarrow$ & $\nearrow$ & $\searrow$ & $\nearrow$ &   & $\searrow$ & $\nearrow$ & $\searrow$ & -0.9399\\
mina-sshd &   &   &   &   &   &   &   &   & $\searrow$ & $\nearrow$ & $\searrow$ & $\searrow$ & $\searrow$ & $\searrow$ & $\searrow$ & $\searrow$ & $\searrow$ & $\nearrow$ & $\searrow$ & -8.2677\\
nifi &   &   &   &   &   &   &   &   &   &   &   &   &   &   & $\searrow$ & $\nearrow$ & $\nearrow$ & $\nearrow$ & $\nearrow$ & 1.8912\\
opennlp &   &   &   &   &   &   &   &   &   &   & $\searrow$ & $\searrow$ & $\nearrow$ & $\searrow$ & $\searrow$ & $\searrow$ & $\searrow$ & $\nearrow$ & $\nearrow$ & 0.1831\\
parquet-mr &   &   &   &   &   &   &   &   &   &   &   &   & $\searrow$ & $\searrow$ & $\searrow$ & $\searrow$ & $\nearrow$ & $\nearrow$ & $\searrow$ & -5.3441\\
pdfbox &   &   &   &   &   &   &   &   & $\searrow$ & $\nearrow$ & $\nearrow$ & $\searrow$ & $\searrow$ & $\searrow$ & $\searrow$ & $\nearrow$ & $\searrow$ & $\nearrow$ & $\searrow$ & -3.1469\\
phoenix &   &   &   &   &   &   &   &   &   &   &   &   &   &   & $\nearrow$ & $\nearrow$ & $\searrow$ & $\nearrow$ & $\nearrow$ & 1.0603\\
ranger &   &   &   &   &   &   &   &   &   &   &   &   &   &   & $\searrow$ & $\nearrow$ & $\searrow$ & $\nearrow$ & $\searrow$ & -3.1562\\
roller &   &   &   &   &   & $\nearrow$ & $\rightarrow$ & $\searrow$ & $\searrow$ &   &   &   & $\searrow$ & $\nearrow$ & $\searrow$ &   &   & $\searrow$ & $\searrow$ & -1.9939\\
santuario-java &   & $\searrow$ & $\nearrow$ & $\searrow$ & $\rightarrow$ & $\searrow$ & $\searrow$ & $\searrow$ & $\searrow$ & $\searrow$ & $\searrow$ & $\searrow$ & $\searrow$ & $\rightarrow$ & $\searrow$ & $\searrow$ & $\searrow$ & $\searrow$ & $\searrow$ & -6.9039\\
storm &   &   &   &   &   &   &   &   &   &   &   & $\searrow$ & $\searrow$ & $\searrow$ & $\nearrow$ & $\searrow$ & $\searrow$ & $\nearrow$ & $\searrow$ & -4.1547\\
streams &   &   &   &   &   &   &   &   &   &   &   &   & $\searrow$ & $\searrow$ & $\nearrow$ & $\searrow$ & $\nearrow$ & $\nearrow$ & $\searrow$ & -12.3819\\
struts &   &   &   &   &   &   & $\searrow$ & $\searrow$ & $\nearrow$ & $\searrow$ & $\searrow$ & $\searrow$ & $\searrow$ & $\searrow$ & $\searrow$ & $\searrow$ & $\searrow$ & $\nearrow$ & $\nearrow$ & 1.2610\\
systemml &   &   &   &   &   &   &   &   &   &   &   & $\searrow$ & $\searrow$ & $\searrow$ &   & $\nearrow$ & $\nearrow$ & $\nearrow$ & $\searrow$ & -4.4465\\
tez &   &   &   &   &   &   &   &   &   &   &   &   &   & $\searrow$ & $\rightarrow$ & $\nearrow$ & $\rightarrow$ & $\nearrow$ & $\searrow$ & -0.3926\\
tika &   &   &   &   &   &   &   & $\searrow$ & $\searrow$ & $\nearrow$ & $\nearrow$ & $\nearrow$ & $\searrow$ & $\searrow$ & $\searrow$ & $\searrow$ & $\searrow$ & $\nearrow$ & $\nearrow$ & 2.5588\\
wss4j &   &   &   &   & $\searrow$ & $\rightarrow$ & $\searrow$ & $\searrow$ & $\searrow$ & $\searrow$ & $\searrow$ & $\nearrow$ & $\searrow$ & $\searrow$ & $\rightarrow$ & $\searrow$ & $\searrow$ & $\searrow$ & $\searrow$ & -8.9378\\
zeppelin &   &   &   &   &   &   &   &   &   &   &   &   &   & $\nearrow$ & $\nearrow$ & $\searrow$ & $\searrow$ & $\nearrow$ & $\searrow$ & -4.4358\\
\hline
mean & $\nearrow$ & $\searrow$ & $\searrow$ & $\nearrow$ & $\nearrow$ & $\searrow$ & $\searrow$ & $\searrow$ & $\searrow$ & $\searrow$ & $\searrow$ & $\searrow$ & $\searrow$ & $\searrow$ & $\searrow$ & $\searrow$ & $\searrow$ & - & - & -3.5035\\
\end{tabular}

\caption{Trends for the number of ASAT warnings over time including sum (\emph{S}), warning density (\emph{R}) and warning density change per year (\emph{R} p.a.)}
\label{tbl:trends_full2}
\end{sidewaystable*}

\subsubsection{\rev{\emph{RQ1.1}: Is the number of ASAT warnings generally declining over time?}}
Table~\ref{tbl:trends_full2} shows, that if we consider the complete lifetime of the project warning density (\emph{R}) increases in only \del{4}\rev{8} of \numberProjects{} projects. The majority of our study subjects improve with regard to warning density.
If we only consider the sum (\emph{S}) the picture is not as clear, here we have more negative trends, i.e., the number of ASAT warnings increase.
This is expected as the number of ASAT warnings usually increases with addition of new code and both are positively correlated as mentioned previously. This shows that if we consider warning density to be a code quality measure, that the code quality steadily increases in most projects.
We also include the value of the slope of the trend for warning density (\emph{R} p.a.) in the table, which indicates a change of warning density in years over the complete lifetime and on average over all projects. We exclude projects where the slope does not met our criteria for F-Score.
The value in column \emph{R} p.a. quantifies the average change in warning density per year, e.g., commons-math removes on average \del{6}\rev{5} ASAT warnings per 1000 Logical Lines of Code per year.
When we consider the mean of all projects we see that on overage \del{3.8}\rev{3.5} ASAT warnings per 1000 LLoC are removed per year.

\begin{center}
\fbox{
\centering
\parbox{0.9\linewidth} {
    \emph{RQ1.1 Summary}: The number of ASAT warnings \rev{from PMD} are not generally declining. However, if we consider warning density then most are declining. Out of \numberProjects{} projects, 43 show declining trends, \del{7}\rev{10} are showing an increasing trend of warning density and \del{4}\rev{1} \del{are}\rev{is} showing minimal changes.
    On average, each project removes \del{3.8}\rev{3.5} ASAT warnings per 1000 LLoC per year.
 }
}
\end{center}

\begin{table}
\centering
\begin{tabular}{l|r|r|r|r}
ASAT group / severity & \rev{M}\emph{R} p.a. & \rev{MED\emph{R} p.a.} & Delta & Remaining\\
\hline
minor & -2.5897 & -1.7125 & 25.74 & 42.95\\
major & -0.8852 & -0.6832 & 7.228 & 12.2\\
critical & -0.1456 & -0.1244 & 1.061 & 2.909\\
\hline
brace rules & -0.7105 & -0.1426 & 6.265 & 4.793\\
design rules & -0.5788 & -0.5187 & 5.507 & 16.36\\
java logging rules & -0.4755 & -0.0902 & 3.094 & 1.968\\
jakarta commons logging rules & -0.4643 & -0.0595 & 1.72 & 3.053\\
naming rules & -0.4221 & -0.2667 & 7.041 & 7.526\\
type resolution rules & -0.3442 & -0.1602 & 2.631 & 3.02\\
controversial rules & -0.2548 & -0.1134 & 1.414 & 6.651\\
optimization rules & -0.2485 & -0.1107 & 2.961 & 3.178\\
basic rules & -0.1301 & -0.0613 & 1.63 & 1.677\\
unnecessary and unused code rules & -0.0705 & -0.0355 & 0.7491 & 0.6812\\
string and stringbuffer rules & -0.0632 & -0.0198 & 0.2866 & 3.263\\
strict exception rules & -0.0562 & -0.0125 & 0.2469 & 3.626\\
security code guideline rules & -0.0401 & 0.0028 & 0.02258 & 0.7211\\
junit rules & -0.0144 & -0.0034 & 0.01282 & 0.004718\\
javabean rules & -0.0079 & -0.0030 & 0.3478 & 0.1154\\
finalizer rules & -0.0006 & -0.0003 & -0.001112 & 0.01842\\
j2ee rules & -0.0003 & -0.0003 & 0.08508 & 0.1346\\
clone implementation rules & 0.0002 & -0.0028 & 0.06261 & 0.08835\\
import statement rules & 0.1722 & -0.0046 & -0.04623 & 1.182\\
\end{tabular}
\caption{\rev{Mean warning density change per year (M\emph{R} p.a.), median (MED\emph{R} p.a.) and delta for ASAT groups and severities}}
\label{tbl:deltas_trends}
\end{table}

\subsubsection{\rev{\emph{RQ1.2}: Which warning types have declined or increased the most over time?}}
\label{sec:rq1.2}
To answer the next research question, we consider how groups of rules have changed in their evolution over the projects lifetime. In this case, we not only report the slope of the trend but also report the delta of the first warning density measurement per project and the last measurement. This provides us with a delta of the absolute number of ASAT warnings per kLLoC per warning group and severity. Due to different project lifetimes we measure these numbers per project and then average the values to end up with a number that encompasses all of our data.

Table~\ref{tbl:deltas_trends} contains all rule groups and severities in our data provided by PMD. It contains the slope of the trend, the average change of warning density per project over the complete lifetime of the project and the remaining warnings per kLLoC.
We can see that, e.g., on average a project removed \del{6.87}\rev{7} naming rule warnings per kLLoC over its complete lifetime and still has \del{7.45}\rev{7.5} warnings per kLLoC left. When considering the trend, we can see that each project, on average, removes \del{0.45}\rev{0.42} naming rule warnings per kLLoC per year.
Moreover, we see that each project on average resolves \del{34.97}\rev{34.03} warnings per kLLoC regardless of its type or severity over its complete lifetime and still has \del{61.18}\rev{58.06} warnings per kLLoC left.

These changes in the number of occurrences of rules by type can hint at potential changes in coding standards in the the years between the beginning of ~2002 and end of 2017. Most prominently brace, design and naming rules, which consist of best practices regarding code blocks and naming conventions, e.g., an \emph{if} should be followed by braces even if it is followed only by a single instruction and class names should be in camel case.
Design rules contain best practices regarding overall code structure, e.g., avoiding deeply nested if statements and simplify boolean returns. We can see that the trend for specialized rules like \rev{Java and} jakarta logging rules is steeper than naming rules, although when we consider the delta it is clear that naming rules are removed far more by number. \rev{Moreover, when we consider the median (MED\emph{R} p.a.) instead of the mean trend (M\emph{R} p.a.) we can see that brace, design, and naming rules also have high median trends.}
A complete list of the rules and their groups as well as their severities is given in the Appendix.

The \del{only}\rev{two} group\rev{s} of warnings that \del{is}\rev{are} increasing by delta, although only slightly, \del{is}\rev{are} finalizer and import statement rules. By \rev{mean} trend only import statement and clone implementation rule violations are increasing slightly.
Finalizer rules are concerned with the correct implementation of finalize() which is called by the garbage collector of Java.
Import statement rules contain rules regarding duplicate imports, unused imports and unnecessary imports, e.g., java.lang or imports of classes from the same package. Clone implementation rules are focused on checking implementations of clone() methods.

\rev{Regarding the severity of the warnings we see that minor severity warnings are resolved the most, major severity warnings second most and critical warnings last.
When we calculate the percentages of reduction in warning density by severity we see that minor and major are reduced by about 37\% each while critical by about 27\%.
This may indicate that developers do not necessarily try to remove all critical warnings. However, this could also be an indication of critical severity warnings being more prone to false positives.}

\rev{The types of warning that declined the most may hint at developer preference or possibly easy resolution of reported warnings. The declining of naming, brace and design rules may also be a consequence of changing coding standards or, more generally, a maturation of Java software coding style. The results may also hint at some rules which are ignored by developers. The density of import statement rules is increasing. This may indicate that this type of rule is more often ignored by developers.}

\begin{center}
\fbox{
\centering
\parbox{0.9\linewidth} {
    \emph{RQ1.2 Summary}: The warning density of Naming, brace and design rules have declined the most. Finalizer \rev{and import statement} rules are the only ASAT warning type\rev{s} that increase.

 }
}
\end{center}

\subsection{\emph{RQ2}: What is the impact of using PMD?}
\label{sec:rq2}

This part of the study discards every commit up until the point in time Maven was introduced as a build system. Although we shorten the project history that is available for analysis, keeping only commits with \rev{a} Maven \del{a} buildfile allows us to be certain that we detect the ASAT inclusion via the Maven configuration
\del{and also}\rev{. Moreover, this} allows us to read custom ruleset definitions and source directories. Utilizing the source directories from the Maven configuration narrows the scope for the files to code only files. We effectively discard tests and tooling which are not part of the build process. Our aim is to be as detailed as possible and counting only the rules that were available at the point in time of the commit\rev{. We also include} \del{and} only \del{the} files that were part of the analysis if the projects developers had run the ASAT via the buildfile.

\begin{table*}
\centering
\tiny
\input{full_table_years_all_overlap_ratio}
\caption{Trends for warning density without overlapping rules (\emph{R}+\emph{o}), green indicates use of PMD in buildfile, red indicates absence of PMD, black indicates partial use of PMD over the year.}
\label{tbl:trends}
\end{table*}

\begin{figure*}
    \includegraphics[width=\textwidth]{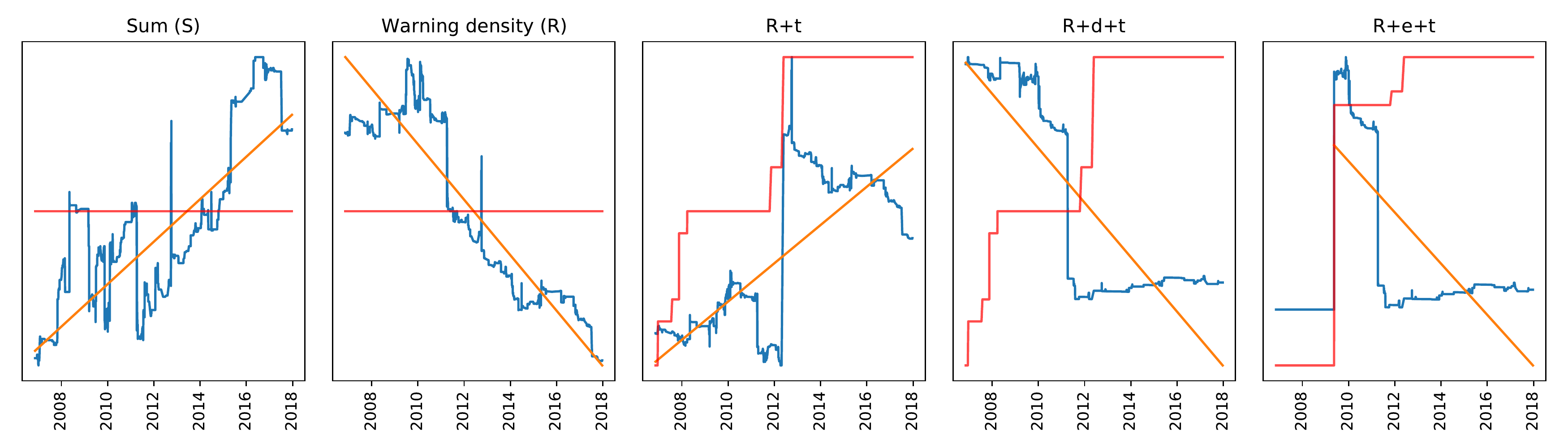}
    \caption{\del{Commons-lang ASAT warning trends}\rev{Example of ASAT warning trends (Commons-lang)}}
    \label{fig:asat_trends_example}
\end{figure*}

\subsubsection{\rev{\emph{RQ2.1}: What is the short term impact of PMD on the number of ASAT warnings?}}
Table~\ref{tbl:trends} shows the trends of ASAT warnings for full years of development. The color indicates if PMD was used for all commits that year: green indicates PMD was used for the complete year, red indicates no use of PMD for the complete year, black indicates partial usage due to introduction or removal of PMD from the buildfile during that year.
\rev{In seven of the 54 projects listed in Table~\ref{tbl:trends}, PMD was removed at least once. We inspected every case to investigate the reasons for the removal.}

\rev{Archiva removed PMD in 2012 when they moved reporting to a parent pom which did not include PMD anymore\footnote{https://repo1.maven.org/maven2/org/apache/archiva/archiva-parent/9/archiva-parent-9.pom}.
Neither the commit message, nor the project documentation mention whether this removal is accidental or not.}

\rev{Commons-bcel briefly introduced and then removed PMD in 2008. The removal does not mention PMD or reports of the build system.
This brief introduction happened at the same time as the move from Ant to Maven as a build system. This indicates that the developers were testing features of Maven.
In 2014 the project included PMD again in its buildfile.
The trend of warnings is declining nonetheless.}

\rev{Commons-compress removed and re-introduced PMD in short order while configuring the build system in multiple commits.
PMD is mentioned explicitly in the commit it was re-added.}

\rev{Commons-dbcp removed PMD in 2014 but that year still shows a declining trend. 
The commit message states that the removal was due to switching to FindBugs. Although the year is not part of this study, PMD was re-added to the build system in 2018.}

\rev{Commons-math changed the ASAT configuration in 2009 - 2011 so that there were at least some commits without an active PMD configuration. Therefore, these are colored black in Table~\ref{tbl:trends}. Those years also had a declining trend. The commit message indicate that in 2009 the reporting section of which PMD is part of was dropped due to a release and later added again.
In 2010 PMD was dropped due to compatibility problems and enabled again in 2011.}

\rev{Commons-validator had some commits in 2008 with PMD enabled. The removal was a conscious decision as it is mentioned in the commit message although the reason is missing.
PMD was added again in 2014.}

\rev{Tika removed PMD in 2009 and never re-introduced it. The commit message mentions removing obsolete reporting from the parent pom.
This indicates that the developers did not act on reported PMD warnings, either because they ignored them completely or because they found that there were too many false positives.}

\rev{We are only considering projects where we can determine the time when PMD was introduced. If it was either introduced together with Maven as a build tool or was introduced before Maven we do not consider it here. We consider projects which introduced PMD and at least used it for a full year afterwards, i.e., a black arrow followed by a green arrow in Table~\ref{tbl:trends}.
The short term impact as estimated by the trend of warning density for 15 projects where PMD was used at least once is declining in 9 projects while 6 projects have an increase in warning trend.
While we expected to see a drop in ASAT warnings after introduction of PMD, this is only the case in 9 of the 15 projects we consider here. An explanation for this result could be that the developers introduce the ASAT but do not immediately scrutinize enough code to make a difference in the short term.} 

\begin{center}
\fbox{
\centering
\parbox{0.9\linewidth} {
    \emph{RQ2.1 Summary}: \rev{The short term impact of PMD on warning density trends is positive in 9 of 15 projects.}
 }
}
\end{center}

\begin{table}
    \centering
    \begin{tabular}{l|c|c|c|c|c}
        Project & \emph{S} & \emph{R} & \emph{R}+\emph{t} & \emph{R}+\emph{d}+\emph{t} & \emph{R}+\emph{e}+\emph{t}\\
        \hline
        
    archiva & $\nearrow$ & $\nearrow$ & $\nearrow$ & $\nearrow$ & $\searrow$\\
calcite & $\nearrow$ & $\nearrow$ & $\nearrow$ & $\nearrow$ & $\rightarrow$\\
cayenne & $\nearrow$ & $\nearrow$ & $\nearrow$ & $\nearrow$ & $\nearrow$\\
commons-bcel & $\nearrow$ & $\searrow$ & $\nearrow$ & $\nearrow$ & $\searrow$\\
commons-beanutils & $\nearrow$ & $\searrow$ & $\nearrow$ & $\nearrow$ & $\rightarrow$\\
commons-codec & $\nearrow$ & $\nearrow$ & $\nearrow$ & $\searrow$ & $\searrow$\\
commons-collections & $\nearrow$ & $\searrow$ & $\nearrow$ & $\searrow$ & $\searrow$\\
commons-compress & $\nearrow$ & $\nearrow$ & $\nearrow$ & $\searrow$ & $\searrow$\\
commons-configuration & $\nearrow$ & $\searrow$ & $\searrow$ & $\searrow$ & $\rightarrow$\\
commons-dbcp & $\nearrow$ & $\searrow$ & $\nearrow$ & $\rightarrow$ & $\rightarrow$\\
commons-digester & $\nearrow$ & $\searrow$ & $\nearrow$ & $\searrow$ & $\searrow$\\
commons-imaging & $\searrow$ & $\searrow$ & $\searrow$ & $\rightarrow$ & $\nearrow$\\
commons-io & $\nearrow$ & $\searrow$ & $\nearrow$ & $\nearrow$ & $\rightarrow$\\
commons-jcs & $\nearrow$ & $\searrow$ & $\nearrow$ & $\searrow$ & $\searrow$\\
commons-jexl & $\nearrow$ & $\searrow$ & $\nearrow$ & $\searrow$ & $\searrow$\\
commons-lang & $\nearrow$ & $\searrow$ & $\nearrow$ & $\searrow$ & $\searrow$\\
commons-math & $\nearrow$ & $\searrow$ & $\nearrow$ & $\searrow$ & $\nearrow$\\
commons-net & $\nearrow$ & $\searrow$ & $\nearrow$ & $\nearrow$ & $\rightarrow$\\
commons-rdf & $\nearrow$ & $\nearrow$ & $\nearrow$ & $\nearrow$ & $\nearrow$\\
commons-scxml & $\nearrow$ & $\searrow$ & $\nearrow$ & $\nearrow$ & $\rightarrow$\\
commons-validator & $\nearrow$ & $\searrow$ & $\nearrow$ & $\searrow$ & $\nearrow$\\
commons-vfs & $\nearrow$ & $\rightarrow$ & $\nearrow$ & $\searrow$ & $\searrow$\\
eagle & $\nearrow$ & $\searrow$ & $\searrow$ & $\nearrow$ & $\rightarrow$\\
falcon & $\nearrow$ & $\searrow$ & $\searrow$ & $\nearrow$ & $\rightarrow$\\
flume & $\nearrow$ & $\searrow$ & $\nearrow$ & $\rightarrow$ & $\rightarrow$\\
giraph & $\nearrow$ & $\searrow$ & $\rightarrow$ & $\nearrow$ & $\rightarrow$\\
gora & $\nearrow$ & $\nearrow$ & $\nearrow$ & $\searrow$ & $\rightarrow$\\
helix & $\nearrow$ & $\nearrow$ & $\nearrow$ & $\rightarrow$ & $\rightarrow$\\
httpcomponents-client & $\nearrow$ & $\searrow$ & $\nearrow$ & $\searrow$ & $\rightarrow$\\
httpcomponents-core & $\nearrow$ & $\nearrow$ & $\nearrow$ & $\nearrow$ & $\rightarrow$\\
jena & $\nearrow$ & $\searrow$ & $\searrow$ & $\searrow$ & $\searrow$\\
jspwiki & $\nearrow$ & $\searrow$ & $\nearrow$ & $\searrow$ & $\rightarrow$\\
knox & $\nearrow$ & $\searrow$ & $\searrow$ & $\nearrow$ & $\rightarrow$\\
kylin & $\nearrow$ & $\searrow$ & $\searrow$ & $\nearrow$ & $\rightarrow$\\
lens & $\nearrow$ & $\searrow$ & $\searrow$ & $\nearrow$ & $\rightarrow$\\
mahout & $\nearrow$ & $\searrow$ & $\nearrow$ & $\searrow$ & $\searrow$\\
manifoldcf & $\nearrow$ & $\searrow$ & $\nearrow$ & $\nearrow$ & $\rightarrow$\\
mina-sshd & $\nearrow$ & $\searrow$ & $\searrow$ & $\nearrow$ & $\rightarrow$\\
nifi & $\nearrow$ & $\nearrow$ & $\nearrow$ & $\nearrow$ & $\rightarrow$\\
opennlp & $\nearrow$ & $\searrow$ & $\searrow$ & $\searrow$ & $\rightarrow$\\
parquet-mr & $\nearrow$ & $\searrow$ & $\searrow$ & $\rightarrow$ & $\rightarrow$\\
pdfbox & $\nearrow$ & $\searrow$ & $\searrow$ & $\searrow$ & $\rightarrow$\\
phoenix & $\nearrow$ & $\nearrow$ & $\nearrow$ & $\nearrow$ & $\rightarrow$\\
ranger & $\nearrow$ & $\searrow$ & $\searrow$ & $\searrow$ & $\nearrow$\\
roller & $\nearrow$ & $\searrow$ & $\nearrow$ & $\searrow$ & $\rightarrow$\\
santuario-java & $\nearrow$ & $\searrow$ & $\searrow$ & $\searrow$ & $\searrow$\\
storm & $\nearrow$ & $\nearrow$ & $\nearrow$ & $\nearrow$ & $\nearrow$\\
streams & $\nearrow$ & $\searrow$ & $\searrow$ & $\nearrow$ & $\rightarrow$\\
struts & $\nearrow$ & $\searrow$ & $\nearrow$ & $\nearrow$ & $\rightarrow$\\
systemml & $\nearrow$ & $\searrow$ & $\searrow$ & $\nearrow$ & $\rightarrow$\\
tez & $\nearrow$ & $\searrow$ & $\nearrow$ & $\nearrow$ & $\rightarrow$\\
tika & $\nearrow$ & $\nearrow$ & $\nearrow$ & $\nearrow$ & $\searrow$\\
wss4j & $\nearrow$ & $\searrow$ & $\nearrow$ & $\searrow$ & $\searrow$\\
zeppelin & $\nearrow$ & $\searrow$ & $\searrow$ & $\nearrow$ & $\rightarrow$\\

    \end{tabular}
\caption{\rev{Trends for the number of ASAT warnings (\emph{S}), warning density (\emph{R}), time-corrected (\emph{R}+\emph{t}), default rules (\emph{R}+\emph{d}+\emph{t}) and effective rules (\emph{R}+\emph{e}+\emph{t}) over all years after Maven introduction.}}
\label{tbl:trends_full}
\end{table}

\subsubsection{\rev{\emph{RQ2.2}: What is the long term impact of PMD on the number of ASAT warnings?}}
Table~\ref{tbl:trends_full} shows the number of ASAT warnings over the projects lifetime from the point in time where Maven was used as a build system, i.e. the point in time where we are sure that we can capture the effective rules of the ASAT.
\emph{S} is the plain sum of the number of warnings, \emph{R} is the warning density. \emph{R}+\emph{t} is the warning density with time-correction where we only count the warnings that PMD supported at the time of the commit. \emph{R}+\emph{d}+\emph{t} is the warning density with only the rules counted that the Maven PMD plugin has enabled by default with time-correction. \emph{R}+\emph{e}+\emph{t} is the warning density with only the rules counted that are definitely enabled via the parsed Maven configuration file, i.e., the most exact and only available in projects where we have the PMD plugin enabled in the Maven configuration file. Figure~\ref{fig:asat_trends_example} visualizes this information \del{for}\rev{using} the project Commons-lang \rev{as an example}. The red line represents the number of rules considered, the blue line is the number of ASAT warnings which is a sum (\emph{S}) in the first subplot and warning density (\emph{R}) in all following subplots. The orange line is the regression line. The number of rules is constant if no time-correction is applied.
Figure~\ref{fig:asat_trends_example} also shows that for the effective ruleset, we only count from the point of inclusion of the ASAT. Otherwise we would skew the data in this case. We should also mention that jumps in effective rules can be due to inclusion of new rules by the developers or by inclusion of new rules for PMD due to group expansion, i.e., the project configures all rules for category A, PMD adds new rules for category A at that point in time which results in rising number of rules considered.

A first interesting result is that if we look at the warning density, there is a downward or neutral trend for all but 13 projects. This means independent of the presence of PMD in the buildfile the overall quality of the code per kLLoC with regards to ASAT warnings improves in most projects. This could be for example through changes in coding style coinciding with some ASAT rules, e.g., no if statement without curly braces. The number of projects is higher than in the previous Section~\ref{sec:rq1} where we considered the complete lifetime of the project. In Section~\ref{sec:rq1}, we observed a rising trend of warning density in only 10 projects.
If we only consider the effective rules (\emph{R}+\emph{e}+\emph{t}) the picture is not that clear, which means even though the developers have the ability to look at the reports containing these warnings the overall quality per LLoC does not always improve. This could be due to perceived or real false positives of the reporting ASAT which are ignored by the developers.

To answer \emph{RQ2.2} we refer to Table~\ref{tbl:trends} again and note the green trends following the introduction of PMD in black. If we add up the slopes of the subsequent years after the introduction of the ASAT, we can estimate the long term impact.
We notice that we have more positive years than negative years in our data following the introduction of an ASAT. Positive years are identified by a decreasing warning density whereas negative years are identified by a increasing warning density.
On a more quantitative note we can sum the slopes of years following the introduction of the ASAT which we report in Table~\ref{tbl:slope_sums}.
We are not listing mina-sshd even though it uses PMD because it was only introduced in 2017 which is the last year of our data, therefore it is excluded from the long term impact analysis.
\begin{table}
    \centering
    \begin{tabular}{l|c}
        Project & \rev{MED\emph{R}+\emph{o} p.a.}\\
        \hline
        archiva & -2.2656\\
cayenne & 0.75534\\
commons-bcel & -15.828\\
commons-beanutils & 0.61942\\
commons-codec & -0.38184\\
commons-collections & -0.65821\\
commons-compress & -1.5768\\
commons-dbcp & -1.7514\\
commons-digester & -0.18101\\
commons-imaging & 0.62411 \\
commons-jcs & -2.3598 \\
commons-jexl & -1.4328 \\
commons-lang & -1.7042 \\
commons-math & -0.96496 \\
commons-rdf & 2.8955 \\
commons-validator & 1.4614 \\
commons-vfs & -0.19471 \\
jena & -0.9757 \\
mahout & -0.51027 \\
range & -5.524\\
santuario-java & -1.5697 \\
stor & -6.735 \\
tika & -14.176 \\
wss4j & -2.5572\\
\hline
Mean & -2.2913\\
    \end{tabular}
    \caption{\rev{Warning density without overlapping rules median change per year (MED\emph{R}+\emph{o} p.a.) after PMD introduction.}}
    \label{tbl:slope_sums}
\end{table}

Table~\ref{tbl:slope_sums} shows the \rev{median} change in warning density per year, e.g., commons-lang decreases the number of warnings per kLLoC by 1.7 per year, which is \rev{almost the same} than its overall decrease over all years (1.8) which can be seen in Table~\ref{tbl:trends_full2}. We can also see that the average change per project is \rev{2.3} which is less than the mean over all projects over all years reported in Table~\ref{tbl:trends_full2} which was \del{3.8}\rev{3.5}.
Nevertheless the projects predominantly show a negative sum which indicates a positive trend in the number of ASAT warnings, i.e., warnings decrease. Only \rev{5} projects, comons-rdf, commons-beanutils, commons-validator, cayenne and \rev{commons-imaging} have a positive sum, i.e., an overall negative trend of warning density after PMD was introduced.
\rev{The long term impact as estimated by the trend of warning density is positive in 19 of 24 projects. On average each project removed 2.3 warnings per kLLoC each year after PMD was introduced in the buildfile.}
Thus we can further conclude that the long term impact of PMD on warning density is better by trend alone than the observed short term impact. Although, its impact is weaker than the overall trend of defect density which encompasses the years where PMD was not present in the buildfile.
\rev{This may be an effect of changing of coding style as the rules that changed the most are related to naming and style (see Section~\ref{sec:rq1.2}).}

\begin{center}
\fbox{
\centering
\parbox{0.9\linewidth} {
    \emph{RQ2.2 Summary}: \rev{The long term impact of PMD on warning density trends is positive in 19 of 24 projects.}
 }
}
\end{center}

\subsubsection{\rev{\emph{RQ2.3}: Does the active usage of custom rules for PMD correlate with higher ASAT warning removal?}}
We first extract all changes to the buildfile and specifically to the custom rules as shown in Table~\ref{tbl:config}. It shows the number of rules changed over the project lifetime and the number of commits where the build file or a configuration file was changed.

\begin{table}
\centering
\begin{tabular}{l|c|c}
    Project & Rule changes & Build changes\\
    \hline
archiva & 45 & 145\\
calcite & 0 & 90\\
cayenne & 130 & 80\\
commons-bcel & 87 & 21\\
commons-beanutils & 45 & 28\\
commons-codec & 94 & 35\\
commons-collections & 45 & 25\\
commons-compress & 133 & 65\\
commons-configuration & 0 & 50\\
commons-dbcp & 45 & 42\\
commons-digester & 45 & 52\\
commons-imaging & 100 & 37\\
commons-io & 0 & 53\\
commons-jcs & 45 & 65\\
commons-jexl & 44 & 96\\
commons-lang & 45 & 73\\
commons-math & 62 & 52\\
commons-net & 0 & 15\\
commons-rdf & 45 & 22\\
commons-scxml & 0 & 31\\
commons-validator & 45 & 25\\
commons-vfs & 45 & 61\\
eagle & 0 & 48\\
falcon & 0 & 57\\
flume & 0 & 40\\
giraph & 0 & 27\\
gora & 0 & 42\\
helix & 0 & 164\\
httpcomponents-client & 0 & 78\\
httpcomponents-core & 0 & 90\\
jena & 45 & 111\\
jspwiki & 0 & 38\\
knox & 0 & 96\\
kylin & 0 & 57\\
lens & 0 & 49\\
mahout & 127 & 261\\
manifoldcf & 0 & 64\\
mina-sshd & 78 & 123\\
nifi & 0 & 133\\
opennlp & 0 & 124\\
parquet-mr & 0 & 60\\
pdfbox & 0 & 61\\
phoenix & 0 & 49\\
ranger & 43 & 70\\
roller & 0 & 10\\
santuario-java & 193 & 79\\
storm & 45 & 17\\
streams & 0 & 76\\
struts & 0 & 155\\
systemml & 0 & 68\\
tez & 0 & 43\\
tika & 45 & 98\\
wss4j & 50 & 96\\
zeppelin & 0 & 77\\

\end{tabular}
\caption{Rule and build changes for each project}
\label{tbl:config}
\end{table}

The number of rule changes are the sum of all deltas of rule changes, this includes additions and removals of rules. 
As we also count the default rules when the ASAT is introduced there is a minimum of 45 rules that are changed if PMD is introduced and there are no custom rules right from the start. If custom rules are added or removed later this number increases.

The true relation to the trends can be seen in Table~\ref{tbl:correlation_changes}, it shows the correlation between the number of rule changes to the general trend of ASAT warnings of the project over each year where the ASAT was used.
The results for this research question could be seen as inevitable because we do not have a lot of rule changes.
Nevertheless, we find that 12 of 25 projects have at least performed some changes to their PMD rulesets.

\rev{While this may sound discouraging to developers, we note that the while the correlation is negligible it retains a negative sign for both correlation measures. This means that while rule changes increase the warning density decreases.}
\rev{However, other factors are probably also important for developers which profit from a well maintained rule set, e.g., acceptance of the ASAT by other developers.}

\begin{table}
\centering
\begin{tabular}{l|c|c}
    Method & Value & P-value\\
    \hline
    Kendall's $\tau$ & -0.17089 & 0.037058\\
Spearman's $\rho$ & -0.20872 & 0.042373\\
\end{tabular}
\caption{Correlation between number of rule changes and warning density}
\label{tbl:correlation_changes}
\end{table}

\begin{center}
\fbox{
\centering
\parbox{0.9\linewidth} {
    \emph{RQ2.3 Summary}: The impact of rule changes, i.e., an active, evolving ASAT configuration on the ASAT warning trends is negligible. Nevertheless, 12 of 25 projects changed their rules at least once.
 }
}
\end{center}

\subsubsection{\rev{\emph{RQ2.4}: Is there a difference in ASAT warning removal trends whether PMD is included in the build process or not?}}
We first have to split our data into two groups, one group contains all years from all projects where PMD was used as indicated by its inclusion in the buildfile, and the other contains all the other years.
We then investigate our two samples for differences\del{, we}\rev{. We} want to know if the two groups have \del{a difference of their}\rev{different} warning trends and if this difference is statistically significant. 
\rev{We now describe the groups of warnings considered here. The \emph{R}+\emph{t} contains the time corrected warning density which contains all possible warnings that were available to the developers at the time of the commit, i.e., a warning added in 2017 would not be included in the warning density of a commit in 2016.
\emph{R}+\emph{d}+\emph{t} contains only the warning density of only the default rules that are enabled by PMD, they are also time corrected. \emph{R}+\emph{e}+\emph{t} contains the time corrected effective rules, this contains the default rules except in cases where developers added custom rule sets. If custom rule sets are found, they are used exclusively. \emph{R}+\emph{e}+\emph{t}+\emph{o} contain the time corrected effective rules without overlapping rules. We subtract PMD warnings which are also reported by other ASATs if the project uses them, we consider FindBugs and Checkstyle.
This removes possible influence in the no PMD group.}

\begin{table}
\centering
\begin{tabular}{l|l|r|r}
Test & Sample & Test Statistic & P-value\\
\hline

Shapiro-Wilk & No PMD & 0.43633 & 1.1615e-20\\
Shapiro-Wilk & PMD & 0.17532 & 6.5589e-31\\
\hline
Levene & Both & 0.52831 & 0.46778\\
\hline
Mann-Whitney-U & Both & 1.6344e+04 & 0.61671\\

\end{tabular}
\caption{Warning density, time corrected rules (\emph{R}+\emph{t}) significance test prerequisites and results}
\label{tbl:all_time_results}
\end{table}

\begin{table}
\centering
\begin{tabular}{l|l|r|r}
Test & Sample & Test Statistic & P-value\\
\hline

Shapiro-Wilk & No PMD & 0.13952 & 9.8708e-25\\
Shapiro-Wilk & PMD & 0.12195 & 1.2032e-31\\
\hline
Levene & Both & 0.34585 & 0.55683\\
\hline
Mann-Whitney-U & Both & 1.4763e+04 & 0.099212\\

\end{tabular}
\caption{Warning density, only default time corrected rules (\emph{R}+\emph{d}+\emph{t}) significance test prerequisites and results}
\label{tbl:default_time_ratio_results}
\end{table}

\begin{table}
\centering
\begin{tabular}{l|l|r|r}
Test & Sample & Test Statistic & P-value\\
\hline

Shapiro-Wilk & No PMD & 0.19915 & 5.1652e-24\\
Shapiro-Wilk & PMD & 0.12195 & 1.2032e-31\\
\hline
Levene & Both & 0.0050551 & 0.94336\\
\hline
Mann-Whitney-U & Both & 1.5212e+04 & 0.20141\\

\end{tabular}
\caption{Warning density, effective time corrected rules (\emph{R}+\emph{e}+\emph{t}) significance test prerequisites and results}
\label{tbl:effective_time_results}
\end{table}

\begin{table}
\centering
\begin{tabular}{l|l|r|r}
Test & Sample & Test Statistic & P-value\\
\hline

Shapiro-Wilk & No PMD & 0.13609 & 5.0062e-24\\
Shapiro-Wilk & PMD & 0.64262 & 2.8805e-20\\
\hline
Levene & Both & 0.4575 & 0.49928\\
\hline
Mann-Whitney-U & Both & 1.2067e+04 & 0.22387\\

\end{tabular}
\caption{Warning density, effective time corrected rules without overlap (\emph{R}+\emph{e}+\emph{t}+\emph{o}) significance test prerequisites and results}
\label{tbl:effective_time_overlap_results}
\end{table}

The Tables~\ref{tbl:all_time_results},~\ref{tbl:default_time_ratio_results},~\ref{tbl:effective_time_results},~\ref{tbl:effective_time_overlap_results} show a difference in the trends of the ASAT warnings between non PMD usage and PMD usage but it is not statistically significant.
Table~\ref{tbl:mwu_reporting} completes the reporting for the Mann-Whitney-U test, it includes sample sizes and the median of the samples. The sample sizes are changing because we remove incomplete years of ASAT usage, overlapping rules and we also remove insignificant trends as described in Section~\ref{sec:approach}.

The reason we found no significant difference could be that the changes resulting from ASAT usage in the buildfile are too small when considering the general number of changes developers apply due to normal code maintenance work. To remove potential influences from overlapping rules of PMD with FindBugs and Checkstyle, we removed them prior to the test in Table~\ref{tbl:effective_time_overlap_results}. This did not change the results significantly.
Overall, we can see \rev{that} the results are not significant, even as we get more detailed, i.e., from just all rules to only the effective rules with removed overlapping rules from other ASATs.

\begin{table}
\centering
\begin{tabular}{l|l|r|r}
    Rules & Samples & Size & Median\\
    \hline
    \multirow{2}{*}{Sum (\emph{S})}& No PMD & 241 & 0.13086\\
& PMD & 136 & 0.02793\\
\hline
\multirow{2}{*}{\emph{R}+\emph{t}}& No PMD & 236 & -0.00093\\
& PMD & 136 & -0.00050\\
\hline
\multirow{2}{*}{\emph{R}+\emph{d}+\emph{t}}& No PMD & 236 & -0.00003\\
& PMD & 136 & -0.00009\\
\hline
\multirow{2}{*}{\emph{R}+\emph{e}+\emph{t}}& No PMD & 236 & -0.00003\\
& PMD & 136 & -0.00002\\
\hline
\multirow{2}{*}{\emph{R}+\emph{e}+\emph{t}+\emph{o}}& No PMD & 200 & -0.00004\\
& PMD & 127 & -0.00002\\
\end{tabular}
\caption{Mann-Whitney-U reporting}
\label{tbl:mwu_reporting}
\end{table}

\begin{center}
\fbox{
\centering
\parbox{0.9\linewidth} {
    \emph{RQ2.4 Summary}: The presence of PMD in the build process has no significant effect on the warning removal trends of warning density.
 }
}
\end{center}

\begin{figure}
    \centering
    \includegraphics[width=0.5\textwidth]{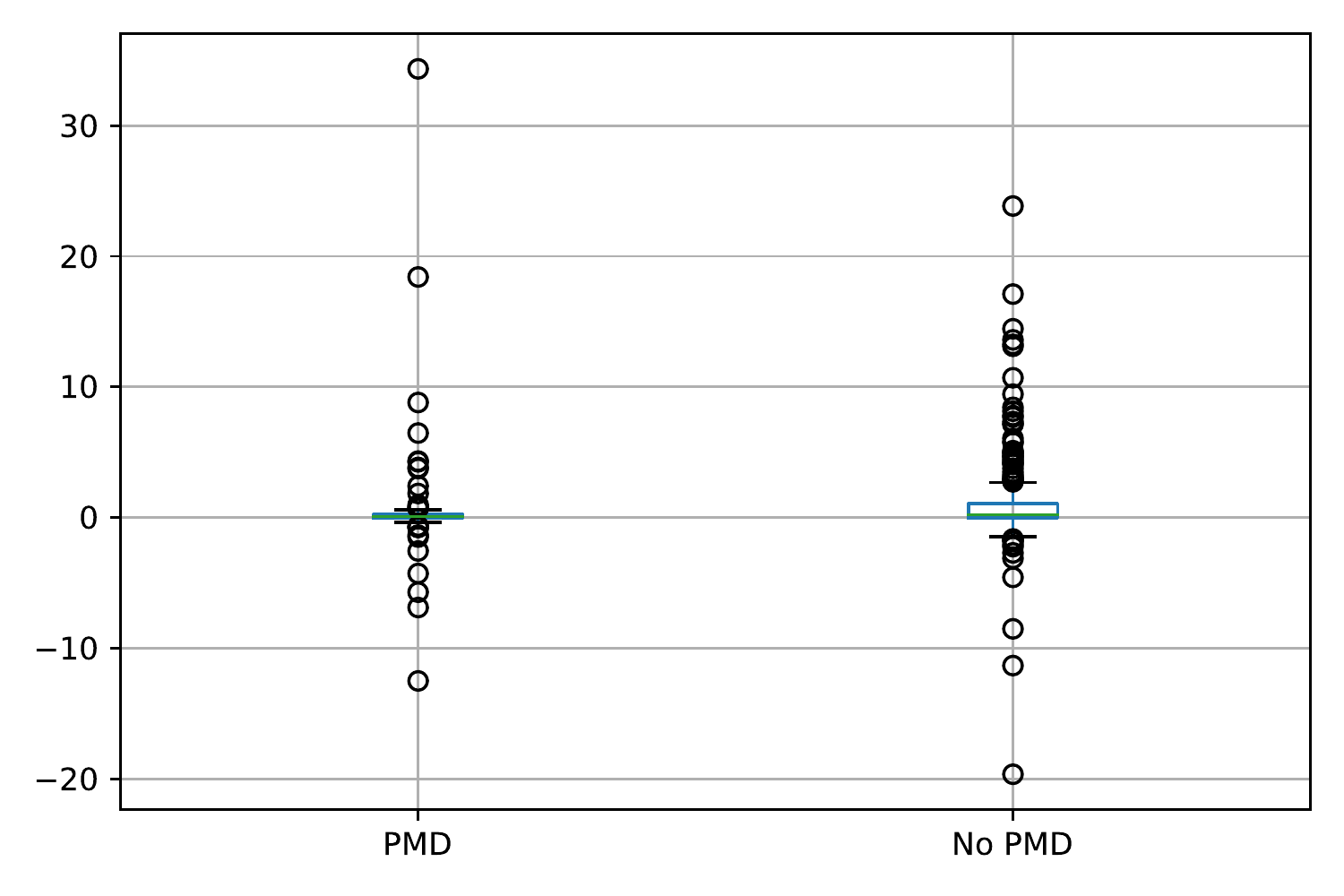}
    \caption{Slope of the sum (\emph{S}) of all ASAT warnings for PMD and non PMD years over all projects.}
    \label{fig:boxplot}
\end{figure}

These results are surprising but looking at our data we can see that there is an effect of PMD usage, just not in a general code quality sense by utilizing the warning density.
If we look at the raw sum of ASAT warnings which we have seen to increase in almost all projects, we can detect an effect. In Figure~\ref{fig:boxplot}, we can see that for most projects the slope of the trend of warning density per year is near 0 whereas for the non PMD using years it is higher.

A possible explanation for this data is, that projects which utilize PMD scrutinize most of the new code, which results in a rising trend of ASAT warnings but only slightly due to some left over warnings or ignored false positive warnings.
In contrast, projects which do not utilize PMD the trend of the sum of ASAT warnings is rising more steeply.

\begin{table}
\centering
\begin{tabular}{l|l|r|r}
Test & Sample & Test Statistic & P-value\\
\hline

Shapiro-Wilk & No PMD & 0.3517 & 5.7853e-22\\
Shapiro-Wilk & PMD & 0.5162 & 2.9655e-25\\
\hline
Levene & Both & 0.97435 & 0.32423\\
\hline
Mann-Whitney-U & Both & 1.2919e+04 & 0.00032041\\
Effect size & Both & 0.17583 & -\\

\end{tabular}
\caption{Sum (\emph{S}) of all ASAT warnings significance test prerequisites and results}
\label{tbl:all_results}
\end{table}

As shown in Table~\ref{tbl:all_results} in case we utilize the sum of all ASAT warnings, the difference is significant, albeit small.

\subsubsection{\rev{\emph{RQ2.5}: Is there a difference in defect density whether PMD is included in the build process or not?}}
We extract issues for all projects from the ITS created in a certain year and then calculate the defect density as described in Section~\ref{sec:defect_density}.
We then build two groups again for years of development where PMD was used and compare it to the second group of years of development where PMD was not used. Instead of the slope of the ASAT warning trends, we now compare the defect densities of the two groups.
We only include years in which PMD was included in the buildfile for every commit or for none to mitigate problems of partial use. The defect density contains only issues marked as a bug by the developers, so we discard improvements, documentation changes.
\rev{Moreover, the issue type used is the one at the end of the data collection. If an issue was misclassified and the classification was changed at some point, the changed classification is the one we use.
This also removes duplicate bug reports from the data, if the developers marked the duplicate bug report as duplicate or invalid as is customary in that case.}

\begin{figure}
    \centering
    \includegraphics[width=0.5\textwidth]{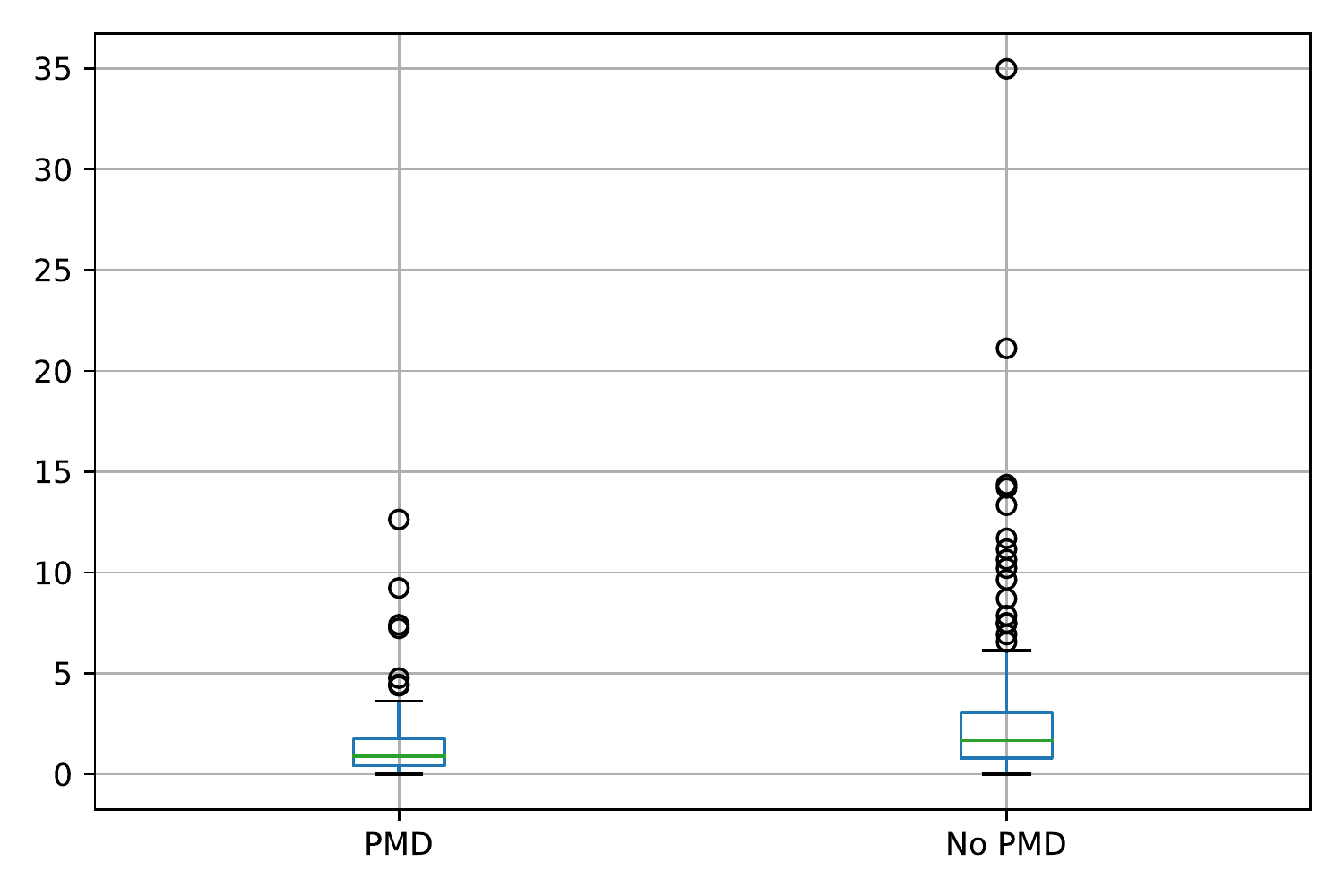}
    \caption{Defect density for PMD and non PMD years over all projects.}
    \label{fig:boxplot2}
\end{figure}

\begin{table}
\centering
\begin{tabular}{l|l|r|r}
Test & Sample & Test Statistic & P-value\\
\hline

Shapiro-Wilk & No PMD & 0.56878 & 1.9853e-24\\
Shapiro-Wilk & PMD & 0.6299 & 9.5835e-17\\
\hline
Levene & Both & 7.6372 & 0.0059968\\
\hline
Mann-Whitney-U & Both & 1.1054e+04 & 7.2559e-08\\
Effect size & Both & 0.26943 & -\\

\end{tabular}
\caption{Defect densities of PMD and non PMD years significance test}
\label{tbl:defect_density}
\end{table}

Figure~\ref{fig:boxplot2} shows the defect densities of the two groups, we can see that PMD using years have a slight advantage of less defect density. Table~\ref{tbl:defect_density} contains the significance test for whether there is a difference between the two groups and its prerequisites.
We can see that years in which PMD is present in the buildfile show a statistically significant difference of defect densities.
To complete the reporting of the Mann-Whitney-U test in Table~\ref{tbl:defect_density}, the sample sizes are 132 and 249 for years where PMD was used and years where PMD was not used. The respective median values are 0.89500 and 1.67743. Resulting in a difference in median of 0.78243 between both samples.
The 95\% confidence interval of the difference in median is (0.36038, 0.86744).

\begin{center}
\fbox{
\centering
\parbox{0.9\linewidth} {
    \emph{RQ2.5 Summary}: Years in which PMD is present in the build process have a lower defect density (about 0.78 less defects per 1000 LLoC in median). The difference is statistically significant, albeit the effect size is small.
 }
}
\end{center}

\rev{This result is very coarse grained. We consider defect density per year which can only hint at a correlation instead of a direct causal relation. However, the ASAT we consider in this study contains a broad set of rules some of which also pertain to more generic maintainability and readability best practices. This may have a more indirect or long term effect on the quality, which is why we decided to include \emph{RQ2.5} in this way.
However, to further validate this result and include confounding factors we build a regularized linear regression model which includes these factors to see if PMD usage still is of importance.
To this end we enhanced the available data with additional features per project per year. We include the number of commits, the number of distinct developers, the year, the number of commits in which PMD was used / not used and the project name as a number.
As a popularity proxy we include the Github information from that project, namely stars and forks.
We train the linear regression model with this data and give the resulting coefficients in Table~\ref{tbl:coeff}.}

\begin{table}
    \centering
    \begin{tabular}{l|r}
        Name & Coefficient\\
        \hline
        Project number & 0.000000\\
        Year & -0.000000\\
        \#commits with PMD & -0.000000\\
        \#forks & 0.049105\\
        \#commits & 0.064325\\
        \#authors & 0.097168\\
        \#stars & 0.106038\\
        \#commits without PMD & 0.217620\\
    \end{tabular}
    \caption{\rev{Linear regression coefficients of the defect density model}}
    \label{tbl:coeff}
\end{table}

\rev{We can see that the regularization of the model removes the project number, and the year as well as the number of commits where PMD was used (although, we note the negative sign). The number of forks, commits, authors, and stars are more important and not removed by the regularization.
The most important feature is the number of commits in which PMD was not used which indicates that it is an important factor when determining defect density.
We also note that except for the number of commits in which PMD was used we retain positive signs on the coefficients of the model. The interpretation is that these factors have a detrimental effect on defect density, i.e., as \#stars or \#commits without PMD increases, so does defect density.}

\section{Discussion}
\label{sec:discussion}
The sum of ASAT warnings is increasing in most of our projects. As the number of ASAT warnings is correlated with the logical lines of code as shown in Table~\ref{tbl:correlation} this is not surprising.
The rising size of the projects is in line with the rules of software evolution~\citep{lehmann} which claim that E-Type software\footnote{Program that performs real world activity, needs to continuously adapt to new requirements and circumstances.} continues to increase in size.
There is no theory for explaining the continued growth of ASAT warnings but it could be interpreted as an indication that some warnings are ignored by developers because they may be deemed unnecessary or false positives (warnings in code without problems). 
An increasing number of ASAT warnings is also supported by the raw data provided by \cite{marcelline} in their replication kit. Although Marcilio et al. investigate real usage of Sonarcube by developers and primarily the resolution times of ASAT warnings, they provide the dates where ASAT warnings are opened and closed.
After transforming their data to show the sum of open ASAT warnings for given days it shows rising sums of ASAT warnings for almost all of the projects.
Furthermore, current research found that only a small fraction of ASAT warnings are fixed by developers~\citep{tse_findbugs,marcelline,warnings_static}.

Table~\ref{tbl:deltas_trends} provide us with additional interesting insights. First of all, code quality, if measured by warning density, is increasing. Second, the different types of ASAT warnings evolve differently.
The order of the ASAT removal trends provided in Table~\ref{tbl:deltas_trends} shows which types of ASAT warnings developers removed most in our candidate projects from 2001-2017. This provides us with a hint of what issues developers deemed most important in that timeframe.
As this first part of our study is independent of ASAT usage, we can not quantify the influence of PMD or other ASATs, i.e., Checkstyle or FindBugs. Nevertheless, the results show that a certain importance is assigned to code readability and maintainability by the developers.
This result is in line with research by \cite{state_static} who found that the majority of actively enabled and disabled rules are maintainability-related.
Beller et al. studied configuration changes. Our work expands on the work of Beller et al. and confirms that not only were the rules more often changed for maintainability related warnings, they were also globally resolved the most.
This finding is also supported by \cite{ci_static} who analyzed CI build logs for ASAT warnings. They found that most builds break because of coding standard violations. However, checking adherence to coding standards via CI quality gates is an industry practice, which is probably a contributing factor.
The most frequently fixed warnings found by \cite{marcelline} also contain rules regarding naming conventions and coding style.
The only groups of ASAT warnings for which we found more introductions than removals in our trend analysis were import \rev{and clone} rules. If we measure by warning density delta between the first and last commit of our analysis period, the only increasing rules are \rev{import and} finalizer rules.
However, as the number of rules here is small and the delta of the change is also small we can not draw any conclusions from this result.

As previously explained we think of code containing less ASAT warnings per line as higher quality code.
In this case study we found that the warning density is decreasing, i.e., the overall quality is increasing.
This is a positive result for the studied open source projects and may also be a positive result for software development in general. 

However, we could not measure a significant difference between the trend of warning density in years of development between PMD usage and no PMD usage.
Although, if we do not consider warning density but the raw sum of warnings there is indeed a measurable, significant difference.
This could be a result of a flattening trend of ASAT warning removal after some time which means the LLoC then becomes the dominating factor of the warning density equation.
This can be seen as evidence of industry best practices like utilizing static analysis tools only on new code as reported by Google~\citep{google_static} and Facebook~\citep{infer_scaling}.
Further evidence of this best practice is shown in the results of short and long term impact of PMD. We found that shortly after the introduction of PMD only 9 of \del{25}\rev{15} projects show decreasing warning density whereas \del{18}\rev{19} of 24 projects show decreasing warning density in the years following PMD introduction.
To the best of our knowledge this would be the first publication to empirically find this effect in open source projects.

As a result of its behavior with regards to the project source code over longer time frames, warning density should be handled with care by researchers including effort aware models using ASAT warnings.
A more targeted warning density as used by \cite{panichella} on the other hand is less problematic. Panichella et al. used warning density targeted on code reviews not on the whole project, therefore avoiding the problem of increasing LLoC on warning density.
Nevertheless, if we want to rank projects by the number of warnings per kLLoC the warning density approach is still viable.

In our ruleset analysis, we found that the number of rule changes does not correlate with the trend of ASAT warnings. Additionally, we found that only a limited number of rule changes in the study subjects are performed. 
This is also supported by \cite{state_static} who found that most configuration files never change. We are now able to expand on the work of Beller et al. and show that there is no direct correlation between ruleset changes and the observable trend of ASAT warnings.

For practitioners, the most interesting result of this study is that defect density, which we use as a proxy for external software quality, is lower when PMD is included in the buildfile. This is also in line with related research from \cite{static_workshop} who found a positive correlation between defects and the number of ASAT warnings by PMD. Although in our case we are not talking about correlations but differences in reported defects. The reported difference in defect density may not necessarily be a result of using PMD and removing its reported warnings but could also be an effect of the developers keeping the codebase healthy, which results in the usage of static analysis tools and subsequently to lower defect densities. Nevertheless, this may serve as a further indication that ASAT warnings and static analysis has a positive impact on software quality evolution.
\rev{Initial results by \cite{Querel2018} show that including static analysis warnings can improve bug prediction models. This is an indication that our results for \emph{RQ2.5} may also hold in a more direct way, i.e., improving direct bug prediction instead of defect density prediction.}

\rev{Our investigation of PMD removal in our study subjects revealed multiple cases where PMD was removed but the removal not explicitly mentioned, e.g., build system reconfigurations, switching parent POMs.
Moreover, some study subjects do not change the PMD default rules. It seems that although some developers advocate static analysis tools like PMD there is no strategy encompassing documentation, continuous integration or integration of project specific rules for local IDEs.
Thus, we recommend adopting a strategy concerning static analysis tools which includes documenting the tools and reasons for inclusion, which rules are enabled for all tools and how the code is checked in different contexts, e.g., continuous integration, code review or local development.}

\rev{Our study revealed a general decrease in warning density. This may be a result of our chosen ASAT as it supports a wide range of rules. Other researchers that focused on security related ASATs come to a different conclusion regarding warning density.
\cite{Penta2009} found that warning density stays roughly constant in their study. More recently, \cite{Aloraini2019} also found similar constant warning density for security related ASATs.
This may indicate that security warnings are harder to find for developers or require specialized knowledge that fewer developers have. In our study, we found that a lot of brace and naming rule related warnings were addressed in our study subjects.
This effect may also have been due to changing or adopting coding standards for Java and may contribute to our finding of declining warning density. However, they were not the only contributing rules to the decline. Our data shows almost every type of rule contributes to the trend of declining warning density.}

\section{Threats to validity}
\label{sec:threats}
In this section, we discuss the threats to validity we identified for our work.
To structure this section we discuss four basic types of validity separately, as suggested by \cite{wohlin}.

\subsection{Construct validity}
Construct validity is concerned with the relation between theory and observation. In our retrospective case study the main source for this threat is due to the observations, i.e., measurements over the course of the change history of our study subjects.
Static analysis warning evolution in test code may be different than in production code, to mitigate this source of noise in our measurements we excluded \rev{all non-production code} for the measurements.
\rev{To validate our exclusion filter we randomly sampled 1\% of the commits of each study subject and the first author manually inspected the changed files for misclassified production files. Out of 3322 production files 3 were misclassified. Out of 1614 non-production files none were misclassified.}

Changes in projects with release branches, e.g., commons-math, may be applied to the release branch as well as the master branch. We mitigate this threat of duplicate measurements by only utilizing a single path through the commit graph.
Considering \emph{RQ2}, the time corrected rules rely on the release date extracted from the PMD changelog. This would in effect mean that as soon as a new PMD version is released the study subjects would be able to see the new rules.
This may not always be a realistic scenario due to delayed updates of, e.g., maven-pmd-plugin. However the data that is available to us does not allow to mitigate this.
The extraction of the effective rules was not possible for every commit due to problems with the Maven buildfile, e.g., XML errors or unavailable parent POMs.
\rev{The buildfile parsing failed for 1361 commits, out of these, 39 errors are due to XML and maven parse errors, 26 errors were due to missing pom.xml files (can happen when the repository is moved but the new folder is not added), 74 due to missing child pom.xml (this happens when the project consists of multiple pom.xml for different modules and the parent references a non existing child) and 1012 errors due to missing parent pom.xml. The last is due to either missing parent pom.xml in the Maven repository or due to a module within the same project not finding its local parent pom.xml. This happens often when a change increases the version of the local parent pom.xml but does not change the referenced parent version in the other modules.}

As there is no way to mitigate this automatically, the rules are assumed to be unchanged for these commits and are changed when the buildfile can be parsed again if there was a change.
To mitigate effects of overlapping static analysis tools, we checked the rules we utilize against the current rulesets of Checkstyle and FindBugs to mark overlapping rules so that we can remove them from the analysis in years and projects where these ASATs are used.
The design of our case study and the chosen statistical tests may influence the results. We include an \del{intensive}\rev{extensive} description of the analysis method and how we preprocess our data prior to the description of the statistical tests we use, and the reasons we chose them.
The statistical tests we utilize in this work depend on their implementation. To mitigate this threat we only rely on well-known and used Python packages scikit-learn~\citep{scikit-learn}, scipy~\citep{scipy} and NetworkX~\citep{networkx}.

\subsection{Internal validity}
Internal validity is threatened by external influences that we did not, or are not able to consider when trying to infer cause-effect relationships.
An external factor we are not able to consider is the usage of tools that are not bound to the Version Control System (VCS), e.g., IDE plugins and cloud services without configurations in the VCS.
This has no impact on questions regarding general trends as in \emph{RQ1} because for this kind of question only the ``end result'', the code that is available in the VCS, is important.
However, for \emph{RQ2} this may interfere with our ability to infer a causal relationship between PMD usage and defect density or warning density.
As the external use of tooling without traces in the VCS is not something that we can include in our available data, we restrict our questions and conclusions to PMD usage via buildfiles and not general usage as in IDE plugins or related tooling.
We are not able to mitigate this effect with our available data and, therefore, note this here as a limitation to our internal validity.

\subsection{External validity}
External validity is concerned with the generalizability of the conclusions we draw in this study.
As we cannot include every Java project, we depend on our sampling of the existing Java projects.
We restricted ourselves to a convenience sample of Java projects managed by the Apache Software Foundation. Nevertheless, our study subjects consist of a diverse set of projects used in different domains to reduce this threat due to the chosen projects.

Furthermore we observe only one ASAT, namely PMD. This restriction is necessary because we cannot rely on all commits in projects being able to compile~\citep{oldversion_nocompile}. Other ASATs, e.g., FindBugs need bytecode files which can be problematic if the project is not being able to compile due to missing dependencies.
Although this is a limitation of our study, PMD includes a wide range of rules. They range from coding style rules to very specific rules concerned with BigInteger usage in Java.

\subsection{Conclusion validity}
Threats to conclusion validity include everything which hinders our ability to draw the correct conclusion about relations between our observed measurements.
For the complete first research question, we are just counting our collected data. Thus, there should be no threat to conclusion validity. We partially plotted and manually verified data to \del{to} validate that our extraction works as expected.
In \emph{RQ2.4} we are comparing the differences in ranks of two samples, i.e., non-PMD and PMD warning trends via a hypothesis test.
The employed hypothesis test, as all hypothesis tests, cannot directly tell us if our assumption is true. We are employing the test under the assumption that there should be a difference and find only a small one. This does not necessarily mean that the difference is really small.
The cause for the difference could also be an effect we do not know about. We tried to mitigate this threat by removing overlapping rules of other ASATs which did not yield significant different results.
For \emph{RQ2.5}, we created two groups for PMD and non PMD using development years, we show that defect density is slightly smaller in years where PMD was used. Although this is what the data shows it could also be a secondary effect not visible to us, e.g., the projects using PMD have a smaller defect density overall due to being more stable feature wise.
For both comparisons of samples we checked the prerequisites for the used statistical test. To correct for the number of statistical tests we employed Bonferroni correction~\citep{bonferroni}.

\section{Conclusion and future work}
\label{sec:summary}
In this work we investigated PMD usage in open source projects in the context of software evolution.
We extracted detailed software repository data over multiple years containing static analysis warnings reported, and extracted additional source code metrics.
In order to determine if our study subjects remove static analysis warnings, we calculated trends of warning density (the number of ASAT warnings per kLLoC) by fitting a linear regression onto cleaned and preprocessed data.
To the best of our knowledge, this is the first longitudinal, commit level study of the evolution of ASAT warnings. Our work complements existing work, which investigated ASAT warnings per warning, by providing a broader, global overview of resolution trends and effects.

To answer our first main research question regarding the evolution of ASAT warnings over time we performed a retrospective case study on a convenience sample of \numberProjects{} open source projects.
We first investigated the evolution of ASAT warnings without taking ASAT inclusion in buildfiles into account.
We found that the general quality of code with regards to ASAT warnings is improving, i.e., warning density is declining. We also found indications of changing coding conventions in our data as the most decreasing types of ASAT warnings were consisting of naming and brace rules.
Moreover, we found that on average every project removes \del{3.8}\rev{3.5} ASAT warnings per thousand LLoC per year.

To answer our second main research question regarding the impact of using PMD on the trend of ASAT warnings we leveraged our evolution data to provide answers to the short and long term effects. 
We found that the short term effects were diverse while the long term effects were positive in the majority of our study subjects.
After that, we split the data into years of development where PMD was included in the build process and years where it was not included. This was done multiple times with different sets of rules. 
We compared both populations and performed a statistical test on both samples. 
The test yielded a surprisingly small difference, i.e., there is no statistical significant difference of using PMD via the build process with regards to the warning density.

We then performed the comparison not on the warning density but on the overall sum of ASAT warnings per commit which is mostly increasing due to its \del{high} correlation with to the size of the projects and possibly false positive warnings.
We found that the difference between years where PMD was used and years where it was not used was significant and that the slopes of ASAT warning trends for years where PMD was used were near zero in most cases.
This could be an indication that best practices that were reported by Google~\citep{google_static} and Facebook~\citep{infer_scaling}, i.e., only new code is scrutinized during static analysis are also utilized in open source projects.

To measure the impact of PMD on software quality, we measured defect density as a proxy metric for external software quality. We compared defect density of samples of development years where PMD was used and where PMD was not used.
We found a statistically significant difference of defect density between years where PMD was used and years where it was not used.
This result shows that for years in which PMD was included in the build process the study subjects had a smaller defect density than in years where PMD was not used.

Future work in this topic, aside from increasing the number of projects in our dataset, could encompass inspecting code changes that increase quality as perceived by developers.
When we determine changes which the developers perceive as quality increasing, we could measure how many ASAT warnings are removed or introduced in these changes.
This would provide a more developer centric viewpoint to complement the defect density view on software quality investigated in this publication.

A subset of data available to us contains manually validated information~\citep{Datensatz}.
This information consists of links between commits and bugs and types of bug reports as well as an improved SZZ~\citep{szz} variant to create links between bug fixes and their inducing changes.
It would be interesting to measure impact of bug fixing changes on the number of ASAT warnings. This could shed light on how many ASAT warnings may be part of a bug.
In our opinion, this would be more within the scope of FindBugs, as PMD contains more general rules. As far as we know there is no\del{t} study that investigated PMD with validated issue types and improved links to inducing changes.

\section*{Acknowledgements}
This work was partly funded by the German Research Foundation (DFG) through the project DEFECTS, grant 402774445.
We also want to thank the GWDG Göttingen\footnote{https://www.gwdg.de}, as without the usage of their HPC-Cluster the data collection would have taken decades.

\bibliographystyle{spbasic}      
\bibliography{literature}

\clearpage

\begin{appendices}
\section{Study subjects}
\begin{center}
  \tiny
  \begin{tabularx}{\textwidth}{l|l|l|l}
  Project & Timeframe & \#Files & \#Commits\\
  \hline
archiva & 2006-2017 & 750 & 7170\\
calcite & 2013-2017 & 1626 & 1543\\
cayenne & 2008-2017 & 3608 & 3655\\
commons-bcel & 2002-2017 & 489 & 1285\\
commons-beanutils & 2002-2017 & 257 & 1028\\
commons-codec & 2004-2017 & 126 & 1562\\
commons-collections & 2002-2017 & 531 & 2885\\
commons-compress & 2004-2017 & 335 & 2197\\
commons-configuration & 2004-2017 & 457 & 2639\\
commons-dbcp & 2002-2017 & 105 & 1578\\
commons-digester & 2002-2017 & 315 & 1143\\
commons-imaging & 2008-2017 & 491 & 981\\
commons-io & 2003-2017 & 234 & 1908\\
commons-jcs & 2003-2017 & 559 & 1288\\
commons-jexl & 2003-2017 & 149 & 1207\\
commons-lang & 2003-2017 & 323 & 4394\\
commons-math & 2004-2017 & 1374 & 5603\\
commons-net & 2003-2017 & 272 & 1570\\
commons-rdf & 2015-2017 & 165 & 221\\
commons-scxml & 2006-2017 & 176 & 760\\
commons-validator & 2003-2017 & 149 & 1233\\
commons-vfs & 2003-2017 & 382 & 1921\\
eagle & 2016-2017 & 1801 & 725\\
falcon & 2012-2017 & 850 & 1669\\
flume & 2012-2017 & 646 & 973\\
giraph & 2010-2017 & 1569 & 876\\
gora & 2011-2017 & 440 & 417\\
helix & 2012-2017 & 823 & 952\\
httpcomponents-client & 2006-2017 & 660 & 2799\\
httpcomponents-core & 2006-2017 & 747 & 2592\\
jena & 2013-2017 & 5669 & 2120\\
jspwiki & 2001-2017 & 529 & 6829\\
knox & 2013-2017 & 1031 & 967\\
kylin & 2015-2017 & 1384 & 2145\\
lens & 2014-2017 & 846 & 763\\
mahout & 2009-2017 & 1220 & 3065\\
manifoldcf & 2011-2017 & 1283 & 1569\\
mina-sshd & 2009-2017 & 931 & 1217\\
nifi & 2015-2017 & 3993 & 2165\\
opennlp & 2011-2017 & 949 & 1703\\
parquet-mr & 2013-2017 & 685 & 501\\
pdfbox & 2009-2017 & 1192 & 6512\\
phoenix & 2015-2017 & 1731 & 1613\\
ranger & 2015-2017 & 944 & 1651\\
roller & 2006-2015 & 610 & 2257\\
santuario-java & 2002-2017 & 660 & 2583\\
storm & 2012-2017 & 1897 & 172\\
streams & 2013-2017 & 528 & 313\\
struts & 2007-2017 & 1953 & 2628\\
systemml & 2012-2017 & 1628 & 3917\\
tez & 2014-2017 & 1089 & 1737\\
tika & 2008-2017 & 988 & 2724\\
wss4j & 2005-2017 & 720 & 2156\\
zeppelin & 2014-2017 & 563 & 2185\\
  \end{tabularx}
\end{center}
\section{PMD rules with groups and severities}

\begin{center}
    \tiny
    \begin{tabularx}{\textwidth}{lll}
        Group & Severity & Rule\\
        Basic Rules & Major & Avoid Branching Statement As Last In Loop\\
Basic Rules & Critical & Avoid Decimal Literals In Big Decimal Constructor\\
Basic Rules & Major & Avoid Multiple Unary Operators\\
Basic Rules & Critical & Avoid Thread Group\\
Basic Rules & Major & Avoid Using Hard Coded IP\\
Basic Rules & Critical & Avoid Using Octal Values\\
Basic Rules & Minor & Big Integer Instantiation\\
Basic Rules & Minor & Boolean Instantiation\\
Basic Rules & Critical & Broken Null Check\\
Basic Rules & Critical & Check Result Set\\
Basic Rules & Critical & Check Skip Result\\
Basic Rules & Critical & Class Cast Exception With To Array\\
Basic Rules & Minor & Collapsible If Statements\\
Basic Rules & Critical & Dont Call Thread Run\\
Basic Rules & Critical & Dont Use Float Type For Loop Indices\\
Basic Rules & Critical & Double Checked Locking\\
Basic Rules & Critical & Empty Catch Block\\
Basic Rules & Minor & Empty Finally Block\\
Basic Rules & Major & Empty If Stmt\\
Basic Rules & Minor & Empty Statement Block\\
Basic Rules & Minor & Empty Statement Not In Loop\\
Basic Rules & Minor & Empty Static Initializer\\
Basic Rules & Major & Empty Switch Statements\\
Basic Rules & Major & Empty Synchronized Block\\
Basic Rules & Major & Empty Try Block\\
Basic Rules & Critical & Empty While Stmt\\
Basic Rules & Minor & Extends Object\\
Basic Rules & Minor & For Loop Should Be While Loop\\
Basic Rules & Critical & Jumbled Incrementer\\
Basic Rules & Critical & Misplaced Null Check\\
Basic Rules & Critical & Override Both Equals And Hashcode\\
Basic Rules & Critical & Return From Finally Block\\
Basic Rules & Major & Unconditional If Statement\\
Basic Rules & Minor & Unnecessary Conversion Temporary\\
Basic Rules & Critical & Unused Null Check In Equals\\
Basic Rules & Critical & Useless Operation On Immutable\\
Basic Rules & Minor & Useless Overriding Method\\
Brace Rules & Minor & For Loops Must Use Braces\\
Brace Rules & Minor & If Else Stmts Must Use Braces\\
Brace Rules & Minor & If Stmts Must Use Braces\\
Brace Rules & Minor & While Loops Must Use Braces\\
Clone Implementation Rules & Major & Clone Throws Clone Not Supported Exception\\
Clone Implementation Rules & Critical & Proper Clone Implementation\\
Controversial Rules & Minor & Assignment In Operand\\
Controversial Rules & Major & Avoid Accessibility Alteration\\
Controversial Rules & Minor & Avoid Prefixing Method Parameters\\
Controversial Rules & Major & Avoid Using Native Code\\
Controversial Rules & Minor & Default Package\\
Controversial Rules & Major & Do Not Call Garbage Collection Explicitly\\
Controversial Rules & Major & Dont Import Sun\\
Controversial Rules & Minor & One Declaration Per Line\\
Controversial Rules & Major & Suspicious Octal Escape\\
Controversial Rules & Minor & Unnecessary Constructor\\
Design Rules & Minor & Abstract Class Without Abstract Method\\
Design Rules & Minor & Abstract Class Without Any Method\\
Design Rules & Critical & Assignment To Non Final Static\\
Design Rules & Minor & Avoid Constants Interface\\
Design Rules & Major & Avoid Instanceof Checks In Catch Clause\\
Design Rules & Minor & Avoid Protected Field In Final Class\\
Design Rules & Minor & Avoid Protected Method In Final Class Not Extending\\
Design Rules & Minor & Avoid Reassigning Parameters\\
Design Rules & Minor & Avoid Synchronized At Method Level\\
Design Rules & Critical & Bad Comparison\\
Design Rules & Minor & Class With Only Private Constructors Should Be Final\\
Design Rules & Critical & Close Resource\\
Design Rules & Critical & Constructor Calls Overridable Method\\
Design Rules & Minor & Default Label Not Last In Switch Stmt\\
Design Rules & Major & Empty Method In Abstract Class Should Be Abstract\\
Design Rules & Critical & Equals Null\\
Design Rules & Minor & Field Declarations Should Be At Start Of Class\\
Design Rules & Minor & Final Field Could Be Static\\
Design Rules & Major & Idempotent Operations\\
Design Rules & Minor & Immutable Field\\
Design Rules & Major & Instantiation To Get Class\\
Design Rules & Minor & Logic Inversion\\
Design Rules & Critical & Missing Break In Switch\\
Design Rules & Minor & Missing Static Method In Non Instantiatable Class\\
Design Rules & Critical & Non Case Label In Switch Statement\\
Design Rules & Critical & Non Static Initializer\\
Design Rules & Critical & Non Thread Safe Singleton\\
Design Rules & Major & Optimizable To Array Call\\
Design Rules & Critical & Position Literals First In Case Insensitive Comparisons\\
Design Rules & Critical & Position Literals First In Comparisons\\
Design Rules & Major & Preserve Stack Trace\\
Design Rules & Major & Return Empty Array Rather Than Null\\
Design Rules & Minor & Simple Date Format Needs Locale\\
Design Rules & Minor & Simplify Boolean Expressions\\
Design Rules & Minor & Simplify Boolean Returns\\
Design Rules & Minor & Simplify Conditional\\
Design Rules & Major & Singular Field\\
Design Rules & Major & Switch Stmts Should Have Default\\
Design Rules & Minor & Too Few Branches For ASwitch Statement\\
Design Rules & Minor & Uncommented Empty Constructor\\
Design Rules & Minor & Uncommented Empty Method\\
Design Rules & Minor & Unnecessary Local Before Return\\
Design Rules & Critical & Unsynchronized Static Date Formatter\\
Design Rules & Major & Use Collection Is Empty\\
Design Rules & Critical & Use Locale With Case Conversions\\
Design Rules & Critical & Use Notify All Instead Of Notify\\
Design Rules & Minor & Use Varargs\\
Finalizer Rules & Major & Avoid Calling Finalize\\
Finalizer Rules & Minor & Empty Finalizer\\
Finalizer Rules & Critical & Finalize Does Not Call Super Finalize\\
Finalizer Rules & Minor & Finalize Only Calls Super Finalize\\
Finalizer Rules & Critical & Finalize Overloaded\\
Finalizer Rules & Critical & Finalize Should Be Protected\\
Import Statement Rules & Minor & Dont Import Java Lang\\
Import Statement Rules & Minor & Duplicate Imports\\
Import Statement Rules & Minor & Import From Same Package\\
Import Statement Rules & Major & Too Many Static Imports\\
Import Statement Rules & Minor & Unnecessary Fully Qualified Name\\
J2EE Rules & Critical & Do Not Call System Exit\\
J2EE Rules & Major & Local Home Naming Convention\\
J2EE Rules & Major & Local Interface Session Naming Convention\\
J2EE Rules & Major & MDBAnd Session Bean Naming Convention\\
J2EE Rules & Major & Remote Interface Naming Convention\\
J2EE Rules & Major & Remote Session Interface Naming Convention\\
J2EE Rules & Critical & Static EJBField Should Be Final\\
JUnit Rules & Minor & JUnit Assertions Should Include Message\\
JUnit Rules & Critical & JUnit Spelling\\
JUnit Rules & Critical & JUnit Static Suite\\
JUnit Rules & Minor & JUnit Test Contains Too Many Asserts\\
JUnit Rules & Major & JUnit Tests Should Include Assert\\
JUnit Rules & Minor & Simplify Boolean Assertion\\
JUnit Rules & Minor & Test Class Without Test Cases\\
JUnit Rules & Minor & Unnecessary Boolean Assertion\\
JUnit Rules & Major & Use Assert Equals Instead Of Assert True\\
JUnit Rules & Minor & Use Assert Null Instead Of Assert True\\
JUnit Rules & Minor & Use Assert Same Instead Of Assert True\\
JUnit Rules & Minor & Use Assert True Instead Of Assert Equals\\
Jakarta Commons Logging Rules & Major & Guard Debug Logging\\
Jakarta Commons Logging Rules & Minor & Guard Log Statement\\
Jakarta Commons Logging Rules & Minor & Proper Logger\\
Jakarta Commons Logging Rules & Major & Use Correct Exception Logging\\
Java Logging Rules & Major & Avoid Print Stack Trace\\
Java Logging Rules & Minor & Guard Log Statement Java Util\\
Java Logging Rules & Minor & Logger Is Not Static Final\\
Java Logging Rules & Major & More Than One Logger\\
Java Logging Rules & Major & System Println\\
JavaBean Rules & Major & Missing Serial Version UID\\
Naming Rules & Minor & Avoid Dollar Signs\\
Naming Rules & Minor & Avoid Field Name Matching Method Name\\
Naming Rules & Minor & Avoid Field Name Matching Type Name\\
Naming Rules & Minor & Boolean Get Method Name\\
Naming Rules & Minor & Class Naming Conventions\\
Naming Rules & Minor & Generics Naming\\
Naming Rules & Minor & Method Naming Conventions\\
Naming Rules & Minor & Method With Same Name As Enclosing Class\\
Naming Rules & Minor & No Package\\
Naming Rules & Minor & Package Case\\
Naming Rules & Minor & Short Class Name\\
Naming Rules & Minor & Short Method Name\\
Naming Rules & Minor & Suspicious Constant Field Name\\
Naming Rules & Critical & Suspicious Equals Method Name\\
Naming Rules & Critical & Suspicious Hashcode Method Name\\
Naming Rules & Minor & Variable Naming Conventions\\
Optimization Rules & Minor & Add Empty String\\
Optimization Rules & Major & Avoid Array Loops\\
Optimization Rules & Minor & Redundant Field Initializer\\
Optimization Rules & Major & Unnecessary Wrapper Object Creation\\
Optimization Rules & Minor & Use Array List Instead Of Vector\\
Optimization Rules & Major & Use Arrays As List\\
Optimization Rules & Major & Use String Buffer For String Appends\\
Security Code Guideline Rules & Major & Array Is Stored Directly\\
Security Code Guideline Rules & Major & Method Returns Internal Array\\
Strict Exception Rules & Major & Avoid Catching Generic Exception\\
Strict Exception Rules & Critical & Avoid Catching NPE\\
Strict Exception Rules & Major & Avoid Catching Throwable\\
Strict Exception Rules & Major & Avoid Losing Exception Information\\
Strict Exception Rules & Minor & Avoid Rethrowing Exception\\
Strict Exception Rules & Minor & Avoid Throwing New Instance Of Same Exception\\
Strict Exception Rules & Critical & Avoid Throwing Null Pointer Exception\\
Strict Exception Rules & Major & Avoid Throwing Raw Exception Types\\
Strict Exception Rules & Critical & Do Not Extend Java Lang Error\\
Strict Exception Rules & Critical & Do Not Throw Exception In Finally\\
Strict Exception Rules & Major & Exception As Flow Control\\
String and StringBuffer Rules & Major & Avoid Duplicate Literals\\
String and StringBuffer Rules & Minor & Avoid String Buffer Field\\
String and StringBuffer Rules & Minor & Consecutive Appends Should Reuse\\
String and StringBuffer Rules & Minor & Consecutive Literal Appends\\
String and StringBuffer Rules & Minor & Inefficient String Buffering\\
String and StringBuffer Rules & Critical & String Buffer Instantiation With Char\\
String and StringBuffer Rules & Minor & String Instantiation\\
String and StringBuffer Rules & Minor & String To String\\
String and StringBuffer Rules & Minor & Unnecessary Case Change\\
String and StringBuffer Rules & Critical & Use Equals To Compare Strings\\
Type Resolution Rules & Major & Clone Method Must Implement Cloneable\\
Type Resolution Rules & Major & Loose Coupling\\
Type Resolution Rules & Major & Signature Declare Throws Exception\\
Type Resolution Rules & Minor & Unused Imports\\
Unnecessary and Unused Code Rules & Major & Unused Local Variable\\
Unnecessary and Unused Code Rules & Major & Unused Private Field\\
Unnecessary and Unused Code Rules & Major & Unused Private Method\\

    \end{tabularx}
\end{center}

\section{PMD rules with overlapping ASATs}
\begin{center}
    \tiny
    \begin{tabularx}{\textwidth}{LL}
        Rule & Overlapping\\
        Avoid Branching Statement As Last In Loop & \\
Avoid Decimal Literals In Big Decimal Constructor & FindBugs (DMI\_BIGDECIMAL\_CONSTRUCTED\_FROM\_DOUBLE) \\
Avoid Multiple Unary Operators & \\
Avoid Thread Group & \\
Avoid Using Hard Coded IP & \\
Avoid Using Octal Values & \\
Big Integer Instantiation & \\
Boolean Instantiation & Checkstyle (ExplicitInitialization) FindBugs (DM\_BOOLEAN\_CTOR) \\
Broken Null Check & \\
Check Result Set & \\
Check Skip Result & \\
Class Cast Exception With To Array & \\
Collapsible If Statements & \\
Dont Call Thread Run & FindBugs (RU\_INVOKE\_RUN) \\
Dont Use Float Type For Loop Indices & \\
Double Checked Locking & \\
Empty Catch Block & Checkstyle (EmptyCatchBlock) \\
Empty Finally Block & \\
Empty If Stmt & \\
Empty Statement Block & \\
Empty Statement Not In Loop & \\
Empty Static Initializer & \\
Empty Switch Statements & \\
Empty Synchronized Block & \\
Empty Try Block & \\
Empty While Stmt & \\
Extends Object & \\
For Loop Should Be While Loop & \\
Jumbled Incrementer & \\
Misplaced Null Check & \\
Override Both Equals And Hashcode & Checkstyle (EqualsHashCode) FindBugs (HE\_EQUALS\_NO\_HASHCODE, HE\_HASHCODE\_NO\_EQUALS) \\
Return From Finally Block & \\
Unconditional If Statement & \\
Unnecessary Conversion Temporary & \\
Unused Null Check In Equals & \\
Useless Operation On Immutable & \\
Useless Overriding Method & \\
For Loops Must Use Braces & Checkstyle (NeedBraces) \\
If Else Stmts Must Use Braces & Checkstyle (NeedBraces) \\
If Stmts Must Use Braces & Checkstyle (NeedBraces) \\
While Loops Must Use Braces & Checkstyle (NeedBraces) \\
Clone Throws Clone Not Supported Exception & \\
Proper Clone Implementation & Checkstyle (SuperClone) FindBugs (CN\_IDIOM\_SUPER\_CALL) \\
Assignment In Operand & Checkstyle (InnerAssignment) \\
Avoid Accessibility Alteration & \\
Avoid Prefixing Method Parameters & \\
Avoid Using Native Code & \\
Default Package & \\
Do Not Call Garbage Collection Explicitly & \\
Dont Import Sun & Checkstyle (IllegalImport) \\
One Declaration Per Line & \\
Suspicious Octal Escape & \\
Unnecessary Constructor & \\
Abstract Class Without Abstract Method & \\
Abstract Class Without Any Method & \\
Assignment To Non Final Static & \\
Avoid Constants Interface & Checkstyle (InterfaceIsType) \\
Avoid Instanceof Checks In Catch Clause & \\
Avoid Protected Field In Final Class & \\
Avoid Protected Method In Final Class Not Extending & FindBugs (DLS\_DEAD\_LOCAL\_STORE, DLS\_DEAD\_LOCAL\_STORE\_OF\_NULL) \\
Avoid Reassigning Parameters & Checkstyle (ParameterAssignment) \\
Avoid Synchronized At Method Level & \\
Bad Comparison & \\
Class With Only Private Constructors Should Be Final & Checkstyle (FinalClass) \\
Close Resource & FindBugs (ODR\_OPEN\_DATABASE\_RESOURCE, ODR\_OPEN\_DATABASE\_RESOURCE\_EXCEPTION\_PATH, OS\_OPEN\_STREAM, OS\_OPEN\_STREAM\_EXCEPTION\_PATH, OBL\_UNSATISFIED\_OBLIGATION, OBL\_UNSATISFIED\_OBLIGATION\_EXCEPTION\_EDGE) \\
Constructor Calls Overridable Method & \\
Default Label Not Last In Switch Stmt & Checkstyle (DefaultComesLast) \\
Empty Method In Abstract Class Should Be Abstract & \\
Equals Null & FindBugs (EC\_NULL\_ARG) \\
Field Declarations Should Be At Start Of Class & \\
Final Field Could Be Static & \\
Idempotent Operations & \\
Immutable Field & \\
Instantiation To Get Class & \\
Logic Inversion & \\
Missing Break In Switch & FindBugs (SF\_SWITCH\_FALLTHROUGH) \\
Missing Static Method In Non Instantiatable Class & \\
Non Case Label In Switch Statement & \\
Non Static Initializer & \\
Non Thread Safe Singleton & \\
Optimizable To Array Call & \\
Position Literals First In Case Insensitive Comparisons & \\
Position Literals First In Comparisons & Checkstyle (EqualsAvoidNull) \\
Preserve Stack Trace & \\
Return Empty Array Rather Than Null & FindBugs (PZLA\_PREFER\_ZERO\_LENGTH\_ARRAYS) \\
Simple Date Format Needs Locale & \\
Simplify Boolean Expressions & Checkstyle (SimplifyBooleanExpression) \\
Simplify Boolean Returns & Checkstyle (SimplifyBooleanReturn) \\
Simplify Conditional & \\
Singular Field & \\
Switch Stmts Should Have Default & Checkstyle (MissingSwitchDefault) FindBugs (SF\_SWITCH\_NO\_DEFAULT) \\
Too Few Branches For ASwitch Statement & \\
Uncommented Empty Constructor & \\
Uncommented Empty Method & \\
Unnecessary Local Before Return & \\
Unsynchronized Static Date Formatter & FindBugs (STCAL\_STATIC\_SIMPLE\_DATE\_FORMAT\_INSTANCE, STCAL\_STATIC\_SIMPLE\_DATE\_FORMAT\_INSTANCE) \\
Use Collection Is Empty & \\
Use Locale With Case Conversions & \\
Use Notify All Instead Of Notify & FindBugs (NO\_NOTIFY\_NOT\_NOTIFYALL) \\
Use Varargs & \\
Avoid Calling Finalize & FindBugs (FI\_EXPLICIT\_INVOCATION) \\
Empty Finalizer & FindBugs (FI\_EMPTY) \\
Finalize Does Not Call Super Finalize & Checkstyle (SuperFinalize) FindBugs (FI\_MISSING\_SUPER\_CALL) \\
Finalize Only Calls Super Finalize & \\
Finalize Overloaded & \\
Finalize Should Be Protected & FindBugs (FI\_PUBLIC\_SHOULD\_BE\_PROTECTED) \\
Dont Import Java Lang & Checkstyle (RedundantImport) \\
Duplicate Imports & Checkstyle (RedundantImport) \\
Import From Same Package & Checkstyle (RedundantImport) \\
Too Many Static Imports & \\
Unnecessary Fully Qualified Name & \\
Do Not Call System Exit & \\
Local Home Naming Convention & \\
Local Interface Session Naming Convention & \\
MDBAnd Session Bean Naming Convention & \\
Remote Interface Naming Convention & \\
Remote Session Interface Naming Convention & \\
Static EJBField Should Be Final & \\
JUnit Assertions Should Include Message & \\
JUnit Spelling & \\
JUnit Static Suite & FindBugs (IJU\_BAD\_SUITE\_METHOD, IJU\_SUITE\_NOT\_STATIC) \\
JUnit Test Contains Too Many Asserts & \\
JUnit Tests Should Include Assert & \\
Simplify Boolean Assertion & \\
Test Class Without Test Cases & FindBugs (IJU\_NO\_TESTS) \\
Unnecessary Boolean Assertion & \\
Use Assert Equals Instead Of Assert True & \\
Use Assert Null Instead Of Assert True & \\
Use Assert Same Instead Of Assert True & \\
Use Assert True Instead Of Assert Equals & \\
Guard Debug Logging & \\
Guard Log Statement & \\
Proper Logger & \\
Use Correct Exception Logging & \\
Avoid Print Stack Trace & \\
Guard Log Statement Java Util & \\
Logger Is Not Static Final & \\
More Than One Logger & \\
System Println & \\
Missing Serial Version UID & FindBugs (SE\_NO\_SERIALVERSIONID) \\
Avoid Dollar Signs & \\
Avoid Field Name Matching Method Name & \\
Avoid Field Name Matching Type Name & \\
Boolean Get Method Name & \\
Class Naming Conventions & Checkstyle (TypeName) FindBugs (NM\_CLASS\_NAMING\_CONVENTION) \\
Generics Naming & Checkstyle (ClassTypeParameterName, InterfaceTypeParameterName, MethodTypeParameterName) \\
Method Naming Conventions & Checkstyle (MethodName) FindBugs (NM\_METHOD\_NAMING\_CONVENTION) \\
Method With Same Name As Enclosing Class & Checkstyle (MethodName) FindBugs (NM\_METHOD\_CONSTRUCTOR\_CONFUSION) \\
No Package & Checkstyle (PackageDeclaration) \\
Package Case & \\
Short Class Name & \\
Short Method Name & \\
Suspicious Constant Field Name & \\
Suspicious Equals Method Name & FindBugs (NM\_BAD\_EQUAL) \\
Suspicious Hashcode Method Name & FindBugs (NM\_LCASE\_HASHCODE) \\
Variable Naming Conventions & Checkstyle (StaticVariableName, ParameterName, ParameterName, ParameterName, ParameterName, ConstantName) FindBugs (NM\_FIELD\_NAMING\_CONVENTION) \\
Add Empty String & \\
Avoid Array Loops & \\
Redundant Field Initializer & \\
Unnecessary Wrapper Object Creation & \\
Use Array List Instead Of Vector & \\
Use Arrays As List & \\
Use String Buffer For String Appends & \\
Array Is Stored Directly & \\
Method Returns Internal Array & \\
Avoid Catching Generic Exception & Checkstyle (IllegalCatch) FindBugs (REC\_CATCH\_EXCEPTION) \\
Avoid Catching NPE & \\
Avoid Catching Throwable & Checkstyle (IllegalCatch) \\
Avoid Losing Exception Information & \\
Avoid Rethrowing Exception & \\
Avoid Throwing New Instance Of Same Exception & \\
Avoid Throwing Null Pointer Exception & \\
Avoid Throwing Raw Exception Types & Checkstyle (IllegalThrows) \\
Do Not Extend Java Lang Error & \\
Do Not Throw Exception In Finally & \\
Exception As Flow Control & \\
Avoid Duplicate Literals & FindBugs (HSC\_HUGE\_SHARED\_STRING\_CONSTANT) \\
Avoid String Buffer Field & \\
Consecutive Appends Should Reuse & \\
Consecutive Literal Appends & \\
Inefficient String Buffering & \\
String Buffer Instantiation With Char & \\
String Instantiation & \\
String To String & \\
Unnecessary Case Change & \\
Use Equals To Compare Strings & Checkstyle (StringLiteralEquality) FindBugs (ES\_COMPARING\_PARAMETER\_STRING\_WITH\_EQ, ES\_COMPARING\_PARAMETER\_STRING\_WITH\_EQ) \\
Clone Method Must Implement Cloneable & FindBugs (CN\_IMPLEMENTS\_CLONE\_BUT\_NOT\_CLONEABLE) \\
Loose Coupling & Checkstyle (IllegalType) \\
Signature Declare Throws Exception & \\
Unused Imports & Checkstyle (UnusedImports) \\
Unused Local Variable & Checkstyle (FinalLocalVariable) FindBugs (DLS\_DEAD\_LOCAL\_STORE, DLS\_DEAD\_LOCAL\_STORE\_OF\_NULL) \\
Unused Private Field & FindBugs (UWF\_UNWRITTEN\_PUBLIC\_OR\_PROTECTED\_FIELD, UWF\_UNWRITTEN\_PUBLIC\_OR\_PROTECTED\_FIELD, UWF\_UNWRITTEN\_PUBLIC\_OR\_PROTECTED\_FIELD, UUF\_UNUSED\_FIELD, URF\_UNREAD\_FIELD, UWF\_UNWRITTEN\_FIELD) \\
Unused Private Method & FindBugs (UPM\_UNCALLED\_PRIVATE\_METHOD) \\

    \end{tabularx}
\end{center}

\end{appendices}

\end{document}